\documentclass[longauth]{aa} 

\usepackage{txfonts}
\usepackage{tablefootnote}
\usepackage{graphicx}   
\usepackage{gensymb}
\usepackage{amsmath}    
\usepackage{amssymb}    
\usepackage{newtxtext,newtxmath}

\usepackage{tablefootnote}
\usepackage{float}
\usepackage[section]{placeins}
\usepackage{placeins}
\usepackage{here}
\usepackage{makecell}
\usepackage[flushleft]{threeparttable}
\DeclareRobustCommand{\VAN}[3]{#2}
\let\VANthebibliography\thebibliography
\def\thebibliography{\DeclareRobustCommand{\VAN}[3]{##3}\VANthebibliography}

\makeatletter
\newcommand*{\rom}[1]{\expandafter\@slowromancap\romannumeral #1@}
\makeatother

\begin{document} 

   \title{Statistical analysis of the causes of excess variance in the 21 cm signal power spectra obtained with the Low-Frequency Array}

   \subtitle{}

   \author{H. Gan \inst{1} \thanks{\email{hgan@astro.rug.nl}}\and
        L. V. E. Koopmans \inst{1}\and
        F. G. Mertens \inst{2,1}\and
        M. Mevius \inst{3}\and
        A. R. Offringa \inst{3,1}\and
        B. Ciardi \inst{4}\and
        B. K. Gehlot \inst{1,5}\and
        R. Ghara \inst{6,7}\and
        A. Ghosh \inst{8}\and
        S. K. Giri \inst{9}\and
        I. T. Iliev \inst{10}\and
        G. Mellema \inst{11}\and
        V. N. Pandey, \inst{3,1}\and
        S. Zaroubi \inst{6,1}\fnmsep 
          }

   \institute{Kapteyn Astronomical Institute, University of Groningen, PO Box 800, 9700AV Groningen, The Netherlands
         \and
             LERMA (Laboratoire d'Etudes du Rayonnement et de la Mati\`{e}re en Astrophysique et Atmosph\`{e}res), Observatoire de Paris,\\ PSL Research University, CNRS, Sorbonne Universit\'{e}, F-75014 Paris, France
         \and
            The Netherlands Institute for Radio Astronomy (ASTRON), PO Box 2, 7990AA Dwingeloo, The Netherlands
         \and
            Max-Planck Institute for Astrophysics, Karl-Schwarzschild-Stra{\ss}e 1, D-85748 Garching, Germany
         \and
            School of Earth and Space Exploration, Arizona State University, Tempe, Az 85281, USA
         \and
            Astrophysics Research Center (ARCO), Department of Natural Sciences, The Open University of Israel,\\ 1 University Road, PO Box 808, Ra'anana 4353701, Israel
         \and
            Department of Physics, Technion, Haifa 32000, Israel
         \and
            Department of Physics, Banwarilal Bhalotia College, GT Rd, Ushagram, Asansol, West Bengal 713303, India
         \and
            Institute for Computational Science, University of Zurich, Winterthurerstra{\ss}e 190, CH-8057 Zurich, Switzerland
         \and
            Astronomy Centre, Department of Physics and Astronomy, University of Sussex, Pevensey II Building, Brighton BN1 9QH, UK
         \and
            The Oskar Klein Centre, Department of Astronomy, Stockholm University, AlbaNova, SE-10691 Stockholm, Sweden
                      }

   \date{Received XXX, 2021; accepted XXX, 2022}

 
  \abstract
   {The detection of the 21 cm signal of neutral hydrogen from the Epoch of Reionization (EoR) is challenging due to bright foreground sources, radio frequency interference (RFI), and the ionosphere as well as instrumental effects. Even after correcting for these effects in the calibration step and applying foreground removal techniques, the remaining residuals in the observed 21 cm power spectra are still above the thermal noise, which is referred to as the "excess variance."}
   {We study a number of potential causes of this excess variance based on 13 nights of data obtained with the Low-Frequency Array (LOFAR). }
   {We focused on the impact of gain errors, the sky model, and ionospheric effects on the excess variance by correlating the relevant parameters such as the gain variance over time or frequency, local sidereal time (LST), diffractive scale, and phase structure-function slope with the level of excess variance. }
   {Our analysis shows that the excess variance, at the current level, is neither strongly correlated with gain variance nor the ionospheric parameters. Rather, excess variance has an LST dependence, which is related to the power from the sky. Furthermore, the simulated Stokes I power spectra from bright sources and the excess variance show a similar progression over LST with the minimum power appearing at LST bin 6h to 9h. This LST dependence is also present in sky images of the residual Stokes I of the observations. In very-wide sky images based on one night of observation after direction-dependent calibration, we demonstrate that the extra power comes exactly from the direction of bright and distant sources Cassiopeia A and Cygnus A with the array beam patterns. }
   {These results suggest that the level of excess variance in the 21 cm signal power spectra is related to sky effects and, hence, it depends on LST. In particular, very bright and distant sources such as Cassiopeia A and Cygnus A can dominate the effect. This is in line with earlier studies and offers a path forward toward a solution, since the correlation between the sky-related effects and the excess variance is non-negligible.}


   \maketitle
%
\section{Introduction}
\label{sec:intro}
The Epoch of Reionization (EoR) is a watershed period in the history of the Universe, where neutral hydrogen (H\rom{1}) in the intergalactic medium (IGM) became ionized by ultraviolet radiation from stars and quasars~\citep{2005SSRv..116..625C,FURLANETTO2006181,doi:10.1146/annurev-astro-081309-130936,Pritchard_2012}. Its study is aimed at shedding light on the fundamental processes of the early Universe~\citep{Madau_1997,1999A&A...345..380S,Tozzi_2000}. 

The Thomson scattering optical depth of cosmic microwave background radiation, the Gunn-Peterson troughs of high-redshift quasar spectra, and the Ly-$\alpha$ emitting high-redshift galaxies (LAEs) are indirect observational probes of the EoR. Studies of these phenomena have suggested that reionization took place at redshift $z=6-10$~\citep{2001AJ....122.2850B,2006AJ....131.1203F,2012A&A...538A..66C,2013ApJ...768..196S,Mortlock2016,2016A&A...596A.108P,10.1093/mnras/stw3351,2018ApJ...864..142D,10.1093/mnras/stz230_Greig_2019,10.1093/mnras/stz632_Mason_2019,2020_Planck_2018_results,Planck_2021_hu}. The most direct probe of the EoR, however, is through the 21 cm line of neutral hydrogen~\citep{1958PIRE...46..240F}. The expected redshift range of the EoR corresponds to observational frequencies between about 120 to 200 MHz, requiring low-frequency radio telescopes. There are mainly two types of observational approaches to detecting the 21 cm signal: (1) using radio interferometry to measure the 21 cm spatial brightness-temperature fluctuations~\citep{10.1111/j.1365-2966.2012.21500.x}; (2) using a single receiver to measure the globally averaged 21 cm brightness temperature as a function of frequency~\citep{1999A&A...345..380S}. In this work, we focus on the first approach. 

Radio interferometers such as the GMRT\footnote{Giant Metrewave Radio Telescope, http://gmrt.ncra.tifr.res.in}~\citep{10.1111/j.1365-2966.2011.18208.x,10.1093/mnras/stt753}, LOFAR\footnote{Low-Frequency Array, http://www.lofar.org}~\citep{van_Haarlem_2013_refId0,Patil_2017,2020MNRAS.493.1662M_Mertens_2020}, MWA\footnote{Murchison Widefield Array, http://www.mwatelescope.org}~\citep{bowman_et_al_2013,2019ApJ...884....1B,2019ApJ...887..141L}, and PAPER\footnote{the Donald C. Backer Precision Array for Probing the Epoch of Reionization, http://eor.berkeley.edu}~\citep{Ali_2015,Cheng_2018,2019ApJ...883..133K},  along more sensitive second-generation instruments such as HERA\footnote{Hydrogen Epoch of Reionization Array, http://reionization.org/}~\citep{DeBoer_2017,theheracollaboration2021hera} and  SKA\footnote{the Square Kilometer Array, http://www.skatelescope.org}~\citep{koopmans2015cosmic} are designed to statistically measure the spatial fluctuations of the 21 cm signal. Furthermore, LEDA\footnote{the Large aperture Experiment to detect the Dark Ages, http://www.tauceti.caltech.edu/leda/}~\citep{greenhill2012hi}, PRIZM\footnote{the Probing Radio Intensity at high z from Marion}~\citep{PRIZM2017}, SARAS 3~\citep{Singh_2017,SARAS3_2021}\footnote{the Shaped Antenna measurement of the background RAdio Spectrum}, and EDGES\footnote{Experiment to Detect the Global EoR Signature}~\citep{Bowman_2018} are aimed at measuring the global 21 cm signal. \cite{Cohen_2018} predicted a tight correlation between features of the 21 cm power spectrum and of the global signal and, hence, the two signals potentially complement each other by providing consistency checks.

The detection of the 21 cm signal is very challenging because the observed sky at low radio frequencies is dominated by strong galactic and extra-galactic emissions, which are three to five orders of magnitudes brighter than the 21 cm signal~\citep{1999A&A...345..380S}. Because it is difficult to model and remove these foregrounds from the observed data, some experiments alternatively use the "foreground avoidance" technique~\citep{Kerrigan_2018}, which discards certain Fourier modes that are heavily affected by the foregrounds. However, giving up Fourier modes (typically at low $k$ values) leads to a loss in sensitivity~\citep{10.1093/mnras/stw161}. The LOFAR-EoR Key Science Project (KSP), therefore, does not use the foreground avoidance strategy; rather, it models and removes the foregrounds. In addition, the data are contaminated by "radio frequency interference" (RFI), largely from human activities~\citep{offringa-2012-morph-rfi-algorithm}. Finally, ionospheric and instrumental effects can distort the signal~\citep{2009arXiv0901.3359L_data_model_Labropoulos,10.1093/mnras/stv1594,10.1093/mnras/stw443_Vedantham2016,2016RaSc...51..927M_Mevius_2016,suppression_paper_Mevius_2020}. Therefore, to extract the 21 cm signal from the data, we need to correct for these effects accurately during calibration~\citep{10.1093/mnras/stw1380_Barry_2016,Trott:2016mry,10.1093/mnras/stw2277_patil2016,Patil_2017}. 

The LOFAR-EoR KSP published its first upper limit on the 21 cm signal power spectrum using only 13 hours of data, with observations centered on the north celestial pole (NCP) in the redshift range of z = 7.9-10.6~\citep{Patil_2017}. It was reported that the residual power exceeded the expected thermal noise level (then defined by the Stokes-V power spectrum) by a factor of 2-3, approaching an excess variance of an order of magnitude in the power spectrum. This was coined as "excess variance"~\citep{10.1093/mnras/stw2277_patil2016,Patil_2017,2020MNRAS.493.1662M_Mertens_2020}. A more recent analysis by \citet{2020MNRAS.493.1662M_Mertens_2020} reported a considerably lower excess variance following improvements in gain calibration, showing that gain fluctuations are unlikely to be the dominant cause of the remaining excess variance. However, the excess variance in the 21 cm power spectra is still significantly larger than the thermal noise level (then defined via time differencing of visibilities) despite the application of very smooth gain solutions. 

In order to analyze more nights of data and separate the 21 cm signal from a mixture of foregrounds as well as instrumental, ionospheric, and RFI effects, it is important to first identify possible causes of excess variance and quantify how much each of them contributes to its level. \cite{2020MNRAS.493.1662M_Mertens_2020} summarized various plausible sources of excess variance: un-subtracted foreground sources that are far from the NCP: polarization leakage~\citep{Jelic_2015,10.1093/mnras/stv1107_Asad_2015,10.1093/mnras/stw1863_Asad_2016}; direction-dependent gain errors due to the overfitting of data and the removal of short baselines ($< 250 \lambda$)~\citep{Patil_2017,suppression_paper_Mevius_2020}; direction-independent gain errors due to the incomplete sky model that are propagated from long baselines to short baselines~\citep{10.1093/mnras/stw1380_Barry_2016,10.1093/mnras/stw2277_patil2016,10.1093/mnras/stx1221_Ewall}; remaining low-level RFI after flagging~\citep{Wilensky_2019,10.1093/mnras/stz175_offringa_2019a}; and uncorrected ionospheric scintillation with very short decorrelation timescales (i.e.,\ of seconds) ~\citep{10.1093/mnras/stw443_Vedantham2016}. Until now, however, no comprehensive study has been carried out to explore whether there is just one dominant contribution from a particular source or whether it is a combination of many factors, and, if so, which factors play a role. 

In this work, we analyze a subset of the above-mentioned possible causes using 13 night observations of the NCP. After the DI and DD calibration, we perform post-calibration flagging to reduce the unfiltered RFI before the DI and DD calibration and filter out bad data. Earlier studies already have shown that early-stage RFI flagging is unlikely to cause effects exceeding the 21 cm signal \citep[][]{10.1093/mnras/stz175_offringa_2019a}. The selection of nights is based on the analysis in~\cite{2020MNRAS.493.1662M_Mertens_2020}. While~\cite{2020MNRAS.493.1662M_Mertens_2020} discarded nights that are highly correlated to each other or have an unusually behaving ionosphere, in this work, we include these nights in the analysis. We specifically focus on parameters that characterize the possible excess variance causes: (1) the gain variance over time or frequency that quantify gain smoothness; (2) the local sidereal time (LST), which is related to the orientation of the instrument with respect to the sky; (3) the diffractive scale, $r_\text{diff}$, and structure-function slope $\beta$, which give good estimates for the ionospheric condition during an observation. We discuss how these effects could contribute to the excess variance. After having chosen the characteristic parameters, we correlate them with the excess variance in the 21 cm signal power spectra to assess whether the chosen parameters contribute to the excess variance.  

The paper is organized as follows: In Section~\ref{sec:data_processing}, we describe the data processing steps and the resulting data products after each step. In Section~\ref{sec_sources}, we define the excess variance in the context of 21 cm signal power spectra and discuss possible causes of the excess variance in detail. We also introduce the relevant parameters which can characterize each possible cause. The correlations between the excess variance and its possible causes are analyzed in Section~\ref{sec:results}. In Section~\ref{sec:conclusions}, we summarize the results and discuss possible improvements in the future.

\section{Observations and data processing}
\label{sec:data_processing}

\begin{table*}
\caption{List of 13 night observational details used for analysis in this paper. Observation dates are given in UTC and LST, respectively, for the later LST-analysis. Excess variance analysis in this paper includes three extra nights of observations (L80850, L203277 and L254871) that are not part of analysis in ~\protect\cite{2020MNRAS.493.1662M_Mertens_2020}.}
\centering
\begin{threeparttable}
 \begin{tabular}{l c c r r r c} 
 \hline
 \noalign{\smallskip}
 Observation ID & UTC Start & UTC End & LST Start & LST End & Duration & Number of SBs\\
  &  &  & [hour] & [hour] & [hour] & \\ 
 \noalign{\smallskip}
 \hline
 \noalign{\smallskip}
 L80847 & 2012-12-31 15:33:06 & 2013-01-01 07:32:56 & 22.6 & 14.7 & 16.0 & 67\\ 
 L80850* & 2012-12-24 15:30:06 & 2012-12-25 07:29:56 & 22.1 & 14.2 & 16.0 & 67\\
 L86762 & 2013-02-06 17:20:06 & 2013-02-07 06:20:01 & 2.8 & 15.9 & 13.0 & 66\\
 L90490 & 2013-02-11 17:20:06 & 2013-02-12 06:20:01 & 3.2 & 16.2 & 13.0 & 66\\
 L196421 & 2013-12-27 15:48:38 & 2013-12-28 07:18:55 & 22.6 & 14.2 & 15.5 & 67\\ 
 L203277* & 2014-02-17 17:14:20 & 2014-02-18 06:17:20 & 3.5 & 16.5 & 13.0 & 65\\ 
 L205861 & 2014-03-06 17:46:30 & 2014-03-07 05:39:49 & 5.1 & 17.0 & 11.9 & 67\\ 
 L246297** & 2014-10-23 16:46:30 & 2014-10-24 05:48:15 & 19.3 & 8.4 & 13.0 & 57\\ 
 L246309 & 2014-10-16 17:01:41 & 2014-10-17 05:35:34 & 19.1 & 7.7 & 12.6 & 67\\
 L253987 & 2014-12-05 15:44:35 & 2014-12-06 07:02:41 & 21.1 & 12.4 & 15.3 & 67\\
 L254116** & 2014-12-10 15:42:54 & 2014-12-11 07:08:31 & 21.4 & 12.8 & 15.4 & 62\\
 L254865 & 2014-12-23 15:45:36 & 2014-12-24 07:17:54 & 22.3 & 13.9 & 15.5 & 67\\
 L254871* & 2014-12-20 15:44:04 & 2014-12-21 07:16:32 & 22.1 & 13.6 & 15.5 & 66\\ [.8ex] 
 \hline
 \noalign{\smallskip}
\end{tabular}
  \begin{tablenotes}
    \small
    \item[*] L80850 and L254871 are highly correlated and L2023277 is strongly affected by the ionosphere. Therefore, these three nights were not used for the upper limit analysis in \protect\cite{2020MNRAS.493.1662M_Mertens_2020}.  
    \item[**] Typically, each night has 65 to 67 SBs. L246297 and L254116, however, have fewer subbands than other observations (with higher than 5\% difference in the number), 57 and 62, respectively. And having fewer subbands can lead to a difference in simulated power.
  \end{tablenotes}
\end{threeparttable}
\label{table:obsv}
\end{table*}

The data analyzed in this work were collected by the LOFAR High-Band-Antenna (HBA) system~\citep{van_Haarlem_2013_refId0} over 13 nights between 2012 to 2014. The duration of observation ranges from 8 to 12 hours. Detailed information for each night is summarized in Table.~\ref{table:obsv}. We use the same observations that were analyzed by~\cite{2020MNRAS.493.1662M_Mertens_2020}. While they excluded three nights (L80850, L254871, and L203277) that are known to have highly correlated residuals or to suffer more from ionospheric effects, we do include these nights in our analysis because we are interested in the sources of the excess variance. The analysis of these three additional nights could potentially provide clues about the excess causes. The observations are centered on the north celestial pole (NCP), for several reasons: (1) the NCP is observable during the night throughout the entire year from the location of LOFAR and the beam is stable with the sky rotating inside of it; (2) due to its location, the $uv$-tracks for the NCP are circular~\citep{Yatawatta_2013} and Earth-rotation synthesis produces a dense $uv$-coverage during a full track. 

The acquired data are processed on the Dawn HPC cluster~\citep{pandey_dawn_2020} by the following steps as described in detail in~\cite{Patil_2017} and~\cite{2020MNRAS.493.1662M_Mertens_2020}: (1) pre-processing including visibility flagging and averaging; (2) calibration including direction-independent (DI) and direction-dependent (DD) calibration; (3) imaging; (4) residual foreground removal; and (5) power-spectrum analysis. While the pre-processing and DI-calibration steps are similar to the ones adopted by~\cite{Patil_2017}, the DD-calibration, foreground removal and power-spectrum analysis strategies have improved since and follow ~\cite{2020MNRAS.493.1662M_Mertens_2020}. Imperfections in the data processing during each step can introduce excess variance. Here, we briefly summarize each step in the processing and its resulting data products. These data products are used in the subsequent analyses.

{\bf Pre-processing:} RFI-flagging is performed on the raw visibility data which are stored in Measurement Set (MS) files as a function of frequency, time, station and polarization. RFI-flagging is done using \textsc{aoflagger}~\citep{offringa-2012-morph-rfi-algorithm} on the highest data resolution (2\,s and 64 channels per sub-band with the four edge channels flagged due to aliasing from the poly-phase filter, resulting in 183 kHz effective width per sub-band) and on a lower resolution after averaging (2\,s and 15 channels per sub-band), respectively. Known bad stations and baselines are also flagged. This flagging typically results in a loss of $\sim5\%$ of the data~\citep{offringa-2012-morph-rfi-algorithm}. After flagging, the data are further averaged to 2\,s and 3 channels per sub-band. 

\begin{table}
\caption{Differences between DI- and DD-gain calibrations.}

\centering
 \begin{tabular}{l l l} 
 \hline
  \noalign{\smallskip}
 Parameter & DI-calibration & DD-calibration \\ [0.3ex] 
 \hline
  \noalign{\smallskip}
 Clustering directions & 2 & 122 \\
 Model components & 1416 & 28755\\
 Baseline cut & > 50$\lambda$ & > 250$\lambda$ \\
 Solution interval & 10 sec & 2.5-20 min\\
 Gain application & $J_\text{DI}\cdot V_\text{data}$ & $V_\text{data}' - J_\text{DD}\cdot V_\text{model}$ \\
 [.8ex] 
 \hline
  \noalign{\smallskip}
\end{tabular}
\label{table:cal_diff}
\end{table}

{\bf DI- and DD-gain calibration:} The visibilities are affected by the ionosphere, instrument, and strong foreground emission. The calibration step is aimed at correcting for the ionospheric and instrumental effects and removing the dominant bright foreground sources in our sky model with their gain solutions applied\footnote{The sky model currently include neither the diffuse emission nor confusion-noise source.}. We describe the propagation of the radio signal from a source to a pair of antenna elements by the radio-interferometry measurement equation (RIME)~\citep{1996A&AS..117..137H,refId0_rime}. The equation relates the observed complex visibility to the model visibility by multiplying it with a series of $2\times2$ gain matrices, known as Jones matrices, which describe different propagation effects including ionospheric and instrumental signal distortions and instrumental (de)polarization of the signal. The calibration is performed by the code \textsc{sagecal-co}~\citep{6051224_skymodel,10.1093/mnras/stv596_Yatawatta_2015b,2016arXiv160509219Y_Yatawatta}. The resulting gains are complex numbers and are calculated per antenna and direction. The gains have similar dimensions as the observed visibilities, namely, (frequency, time, polarization) and gains from two different antennas can be combined to obtain gain solutions for baselines. 
    
The gain calibration process is done in two steps: direction independent (DI-) and direction dependent (DD-) calibration. The main differences between the two calibration steps are the sky models, baseline cuts, clustering of sky components during calibration and the way those gains are applied (summarized in Table.~\ref{table:cal_diff}). DI-calibration uses only a subset of the NCP sky model due to computational limitations, that is,\ the 1416 brightest components with an apparent-flux limit of $>$35\,mJy, including the relatively bright source 3C61.1 (a total flux about 35 Jy at $150$ MHz,~\cite{refId0/buildsky}). The sky components are clustered into two directions: 3C61.1 and the remainder, each with their gain solution. In DD calibration, the full NCP sky model with 28755 components, including 18 shapelets and Cygnus A and Cassiopeia A (Cas A and Cyg A, hereafter), is used. This sky model is clustered into 122 groups of sources to handle DD gain variations due to the time-varying primary beam and the ionosphere. DI and DD calibration exclude baselines that are shorter than 50$\lambda$ and 250$\lambda$, respectively. The 50$\lambda$ cut is chosen to minimize effects from the Galactic diffuse emission and the 250$\lambda$ cut is chosen to avoid signal loss (for details, see~\cite{Patil_2017,2020MNRAS.493.1662M_Mertens_2020}). In addition, gain solutions are applied differently in DI and DD calibration: DI gain solutions are directly applied to the observed visibilities, whereas DD gain solutions are applied to the sky-model visibilities and their product is subtracted from the observed visibilities.  
    
{\bf Imaging:} The residual visibilities after DI and DD calibration are gridded and imaged using \textsc{wsclean}~\citep{2014MNRAS.444..606O} for each sub-band and two polarizations (Stokes I and V) to create an ($l$,$m$,$\nu$) image cube. Each image in units of Jy/PSF is naturally weighted and has a field of view (FoV) limited to $12^{\circ}\times12^{\circ}$ with 0.5 arcmin pixel size. Images from different sub-bands are combined to form an image cube. After gridding, these images are trimmed with a Tukey spatial filter of $4^{\circ}\times4^{\circ}$ FoV to focus on the primary beam and maximize the sensitivity. For the thermal-noise estimation, we additionally create alternating (i.e.,\ "even" and "odd") 10 second images to generate gridded and  time-differenced visibilities~\citep{2020MNRAS.493.1662M_Mertens_2020}. Any effect that changes on a timescale that is much longer than 10 seconds will therefore cancel in this difference and we consider this an excellent estimator of the thermal noise, in agreement with calculations based on the SEFD.
    
{\bf Residual foreground removal:} After DD calibration and subtraction of the gain-corrected sky-model visibilities, the residual visibilities still contain (partially polarized) emission from the diffuse foregrounds and sources that are not included in the calibration model. These residuals are also only corrected for DI gain errors. Imaging is performed with settings that limit gridding errors to a negligible error~\citep{2019A&A...631A..12O_Offringa_2019b}. Gaussian Process Regression (GPR) \citep{10.1093/mnras/sty1207/gpr,2020MNRAS.493.1662M_Mertens_2020} is subsequently applied to remove residual foregrounds from the inner $3^\circ \times3^\circ$ of the images, restricted to well within the first null of the primary beam and used in power-spectrum estimates. GPR uses the property that the Stokes-I components of foregrounds have spectrally-smooth emission. This can be distinguished from the 21 cm signal, which is rapidly fluctuating in frequency \citep{10.1111/j.1365-2966.2006.10502.x,10.1111/j.1365-2966.2006.10919.x}. 
    
{\bf Power-spectrum analysis:} After removing the residual foregrounds, various power spectra are estimated by taking the Fourier transform of visibility cubes along the frequency axis. After the Fourier transform, the resulting delays and baseline lengths are transformed to wave numbers in units of inverse co-moving distance ($\text{Mpc}$) following the cosmological convention of \cite{Morales_2004}. Finally, the power spectra $P(k)$ are transformed into units of $\text{K}^2h^{-3}c\text{Mpc}^3$. 

After processing 13 night observations from LOFAR, following the described pipeline and given the sensitivity of the instrument, we should expect that final residual power spectra are dominated by thermal noise and are relatively flat as a function of $k_{||}$. However, the resulting power spectra are significantly higher than the thermal noise level and we discuss possible causes in the next sections.

\section{Thermal noise and excess variance}
\label{sec_sources}
\begin{figure*}
    \centering
    \includegraphics[width=16cm]{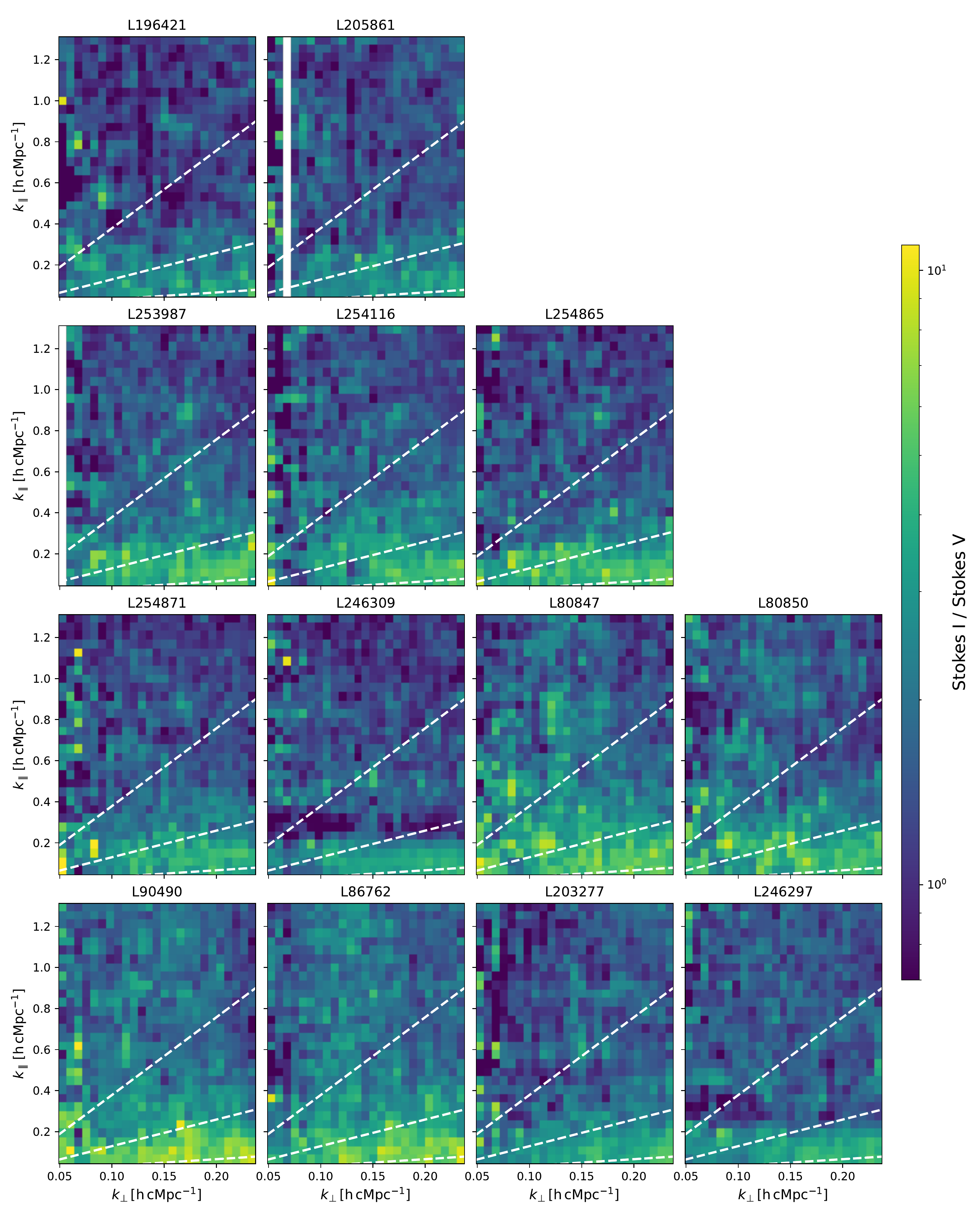}
    \caption{Excess variance power spectra ratios (i.e.,\ the Stokes I over Stokes V ratio) for the 13 nights of LOFAR observations after DI and DD calibration and foreground removal (sky model and GPR). The dashed lines indicate, from bottom to top, the 5\degr~(the primary beam), 20\degr~, and instrumental horizon delay lines. The excess variance is larger in the wedge-like region under the dashed lines. Some baselines are flagged in L205852 ($k_\perp\sim0.07 \text{ }hc\text{Mpc}^{-1}$) and in L253987 ($k_\perp\sim0.05 \text{ }hc\text{Mpc}^{-1}$).}
    \label{fig:full_night_excess}
\end{figure*}

The raw data contain various components: foreground emission, RFI, instrumental and ionospheric distortions, thermal noise, and the 21 cm signal. If DI and DD calibration and foreground removal are applied correctly, the data after the processing, including calibration and foreground removal, should only contain the 21 cm signal and thermal noise. However, the data processing, especially calibration and foreground removal, is not perfect. These imperfections in the processing could result in adding extra power ("excess variance") to the final data products. Fig.\ref{fig:full_night_excess} shows the excess variance of the 13 nights of LOFAR-HBA NCP observations after calibration and foreground removal (i.e.,\ sky-model and GPR subtractions). Although we do not completely rule out the scenario that this excess power is some real power from the sky, or a rapid  gain variation that is not yet understood, our current analysis suggests that this excess is mostly artificial and it should be distinguishable from thermal noise. Thermal noise is incoherent and, hence, it can be reduced by adding more visibilities. On the other hand, part of the excess variance is coherent \citep{2020MNRAS.493.1662M_Mertens_2020}. The coherent part of excess variance does not average down by increasing the number of visibilities contrary to the thermal noise~\citep[see][]{10.1093/mnras/stw2277_patil2016,Patil_2017,2020MNRAS.493.1662M_Mertens_2020}. 

Below, we discuss plausible causes for the excess variance in the LOFAR data and its power spectra and discuss its characteristics in more detail. We provide parameters that can be used to describe the correlation between the excess variance and its causes.

\subsection{Thermal noise}

\cite{Patil_2017} used the Stokes-V power spectra as an estimator of the thermal noise contribution to the 21 cm power spectrum, based on the expected low level of circularly polarized emission from radio sources. Given the fact that thermal noise is still expected to be much higher than the 21 cm signal, ideally, the ratio between the residual Stokes I and Stokes V power spectra should approach unity from above. These ratios for the three redshift bins considered in~\cite{Patil_2017}, however, showed an excess well beyond thermal noise, typically by a factor 2-3. Furthermore, \cite{2020MNRAS.493.1662M_Mertens_2020} found that the Stokes-V signal is not fully dominated by thermal noise, even after improved calibration. For example, some Stokes I emission can instrumentally leak into Stokes V, and random errors on the gain solutions can increase the power in Stokes V above thermal noise, via such leakage. 

To obtain a much more accurate thermal noise estimation and to account for any possible leakage,~\cite{2020MNRAS.493.1662M_Mertens_2020} used two additional thermal noise metrics: the time-difference visibilities $\delta_\text{t}V(u,v,\nu)$ and the sub-band-difference visibilities $\delta_{\nu}V(u,v,\nu)$. The time-difference visibilities are calculated by subtracting two alternating ("even" and "odd") gridded visibility sets with 10 second time differences. These time-differenced visibility sets should produce an excellent thermal noise estimate because most of the foreground and some ionospheric errors cancel out on this timescale, even though it has been found that on very short baselines, the timescale is comparable to the decorrelation timescale of ionospheric scintillation noise~\citep{10.1093/mnras/stv1594}. The second estimator is calculated by subtracting visibility sets between 195.3-kHz sub-bands. Because the subtraction between sub-bands removes spectral structures on the sub-band or larger scale, this estimator is more sensitive to the spectrally-uncorrelated excess noise in the data.

In addition, the excess variance in residual power spectra appears to have two components \citep{2020MNRAS.493.1662M_Mertens_2020}: (1) spectrally uncorrelated excess on short baselines ($<100\lambda$) and (2) spectrally and temporally (between a few nights) correlated excess ($l_\text{ex}\sim0.25-0.45$ MHz) that is stronger in the foreground-dominated wedge. The latter is currently not removed by the GPR method, which separates different signal components by using distinct frequency-correlation characteristics in each component~\citep{10.1093/mnras/sty1207/gpr}. The first excess component does not have specific spectral coherence, so it is difficult to remove via GPR. On the other hand, the frequency coherence scale of the second excess component is similar to that of the 21 cm signal. This similarity makes it difficult to distinguish them only via GPR. 

Thanks to huge improvements in the calibration process compared to the previous work by~\cite{Patil_2017}, the excess variance~\citep{2020MNRAS.493.1662M_Mertens_2020,suppression_paper_Mevius_2020} is significantly reduced and the noise level almost reaches thermal noise in the region $k_\parallel>1 \text{ }hc\text{Mpc}^{-1}$. However, the excess variance still remains higher than thermal noise, especially in the region of $k_\parallel\leq1 \text{ }hc\text{Mpc}^{-1}$. 

\subsection{Excess variance}\label{subsec:possible_causes}

In this subsection, we briefly review how various effects can contribute to the observed excess variance. 

\subsubsection{Radio frequency interference and data flagging} 

The frequency range of 134.1782-146.0922 MHz, included in our observations, is contaminated by RFI. Currently,~\textsc{aoflagger} flags RFI on a $\sim$12.2 kHz scale. As a result, a small number of channels and sub-bands are fully flagged after the pre-processing. Low-level RFI under the threshold can also affect the data and introduce frequency structures at high $k_{\parallel}$. Additionally, ~\cite{10.1093/mnras/stz175_offringa_2019a,Wilensky_2020} found that a combination of data missing due to RFI flagging and averaging data could result in "excess power" at high $k_\parallel$. However, given the current approach to RFI flagging, these effects in our data should be significantly smaller than the predicted 21 cm signal and thermal noise~\citep{10.1093/mnras/stz175_offringa_2019a}. 

In this paper, therefore, we do not investigate residual RFI further, since there is currently no clear evidence for its contribution to the observed excess variance. Flagging data, however, can have an effect, since it can create an achromatic point spread function. This will be investigated in more detail in a future study.

\subsubsection{Ionospheric phase error and scintillation}
\label{par:ionosphere}

Ionospheric propagation effects can be an important source of errors at low radio frequencies~. They depend on multiple factors, such as\ time, frequency, direction, and baseline \citep{1456582_review_morphology_ionosphere_Aarons_1982,2001isra.book.....T}. %
During the calibration step, some of these ionospheric effects can be corrected for~\citep{1984ARA&A..22...97Pearson,refId0_rime}, but not all of them, especially, on shorter baselines where effects vary rapidly with time and direction \cite[see][for an extensive discussion]{10.1093/mnras/stv1594}. The DI calibration is done at 10 second intervals and only corrects for a flux-weighted ionospheric phase variation over the entire field of view. In DD-calibration, our sky model is clustered into 122 directions and any DD ionospheric phase fluctuations on timescales that exceed the solution interval (typically 2.5-20 minutes) are solved for each direction. Hence, the directional and temporal dependence of the ionosphere can only be accounted for in a limited way. 
According to~\cite{10.1093/mnras/stv1594}, under typical ionospheric conditions -- at a frequency of $\nu=150$ MHz, height $h=300$ km and ionospheric turbulence travelling along a bulk of wind at speed of $v=100-500$ km/hr -- the decorrelation time for short baselines ($<300$ m) varies between 4 and 22 seconds, while the decorrelation time for long baselines ($>2$ km) ranges between 30 and 150 seconds. Thus, the typical calibration time intervals are longer than the actual decorrelation timescales of ionospheric phase errors (about a few seconds) on short baselines. This results in gain errors and leaves rapidly varying ionospheric scintillation in the data. The remaining scintillation (known as "scintillation noise") may appear as frequency correlated excess variance in power spectra~\citep{Koopmans_2010,10.1093/mnras/stv1594,10.1093/mnras/stw443_Vedantham2016}. This type of excess variance is contained in the foreground wedge~\citep{10.1093/mnras/stv1594}. Similarly, gain errors induced by the ionosphere on intermediate and longer baselines are also applied to the sky model on short baselines and thus can lead to another indirect excess variance (see Section~\ref{subsub:gain_cal} for more details on the same effect due to an incomplete or incorrect sky model) that can also enter the EoR window \citep[e.g.,][]{10.1093/mnras/stx1221_Ewall}.  

In addition to phase errors, the "Faraday rotation" causes any linearly polarized signal to rotate, including any linear polarization leaking to Stokes I due to instrumental polarization leakage \citep[][]{Jelic_2010} and it is currently considered to be a second-order effect. This effect can contaminate "the EoR window," but is expected to be far below the current excess variance seen in our LOFAR data~\citep{10.1093/mnras/sty258_Asad_2018}.

The ionospheric phase variance per baseline to first-order follows a power law, namely, the phase structure function~\citep{Tol_2009, 2016RaSc...51..927M_Mevius_2016}, where we can obtain two metrics for the ionospheric quality representation, the fitted slope $\beta ,$ and the diffractive scale $r_\text{diff}$. The exponent of the structure function $\beta$ varies from 5/3 to 2, depending on the ionospheric structure. A lower power index of 5/3 comes from Kolmogorov turbulence, while the higher index 2 suggests non-turbulent structures such as traveling ionospheric disturbances (TIDs)~\citep{053a1363370540658ce7bd7cc38617ad_van_Velthoven} or density ducts~\citep{https://doi.org/10.1002/2015RS005711_loi}.

The diffractive scale is the length scale over which the phase variance corresponds to 1 rad$^2$. A large diffractive scale indicates longer time coherence of the signal and smaller phase fluctuations over a given field of view. Therefore, the diffractive scale should be a good first-order representation of the ionospheric state during an observation.
We correlate $\beta$ and $r_\text{diff}$, respectively, with the excess variance to assess whether there is any correlation between the ionospheric quality and the excess variance in power spectra.

\subsubsection{Gain-calibration errors}
\label{subsub:gain_cal}
The calibration of LOFAR EoR KSP data is completed in two steps (see Section~\ref{sec:data_processing}): DI- and DD-calibration. Gain solutions in the two steps are obtained with different (although motivated) choices for the sky models, baseline cuts, time, and frequency resolution. The obtained solutions are applied to our data differently. While DI gains are directly applied to the visibility data, DD gains are applied first to the model visibilities and subsequently subtracted from DI-calibrated visibility data. In this subsection, we discuss how gain errors in DI- and DD-calibration (summarized in Table.~\ref{table:cal_diff}) from different sources can contribute to the excess variance.

\paragraph*{Sky-model incompleteness: }
Currently, we use a sky model consisting of 28,773 unpolarized components (28,755 delta function components and 18 shapelets~\citep{6051224_skymodel,2020MNRAS.493.1662M_Mertens_2020}) for DD calibration. For DI calibration, only the 1416 brightest components of the sky model are used. This subset is chosen from the NCP model with an apparent flux limit of $>$35 mJy to reduce the processing time but still fulfil the signal-to-noise (S/N) ratio required for the DI calibration at high time resolution~\citep{2020MNRAS.493.1662M_Mertens_2020}. The sky model is built from wide-field images with $\sim6$ $\text{arcsec}$ resolution by an iterative process using a program called \textsc{buildsky}~\citep{Yatawatta_2013} and all Dutch baselines. The source components in the model have individual spectral dependencies. The model, even though it is already quite extensive, has some known limitations: diffuse emission from the Milky Way is not part of the sky model (even though it is not negligible on baselines < 100$\lambda$); extra-galactic sources typically have an angular size similar to the resolution of LOFAR images ($\sim$6 arcsec when using all Dutch baselines $\sim100$ km), which limits the spatial accuracy of the model for each source.

These limitations make the construction of an accurate sky model difficult~\citep{Patil_2017}. The difference between the true sky and the modelled sky, however, is less than $5\%$ on the flux scale according to~\cite{10.1111/j.1745-3933.2012.01251.x,Patil_2017,2020MNRAS.493.1662M_Mertens_2020} \footnote{The intensity scale of the model is set by the compact source NVSS J011732+892848 ($\text{RA}=1^{\text{h}} 17^{\text{m}} 33^\text{s}$, ~$\text{Dec.}~=89^\circ 28\text{'} 49.4\text{''}$ ~in J2000) with an intrinsic flux of 8.1 Jy with $5\%$ accuracy~\citep{2020MNRAS.493.1662M_Mertens_2020}. }. This flux difference is distributed over many sources and acts as a "confusion noise source," limited to the foreground wedge of 21 cm power spectra. It does not change the absolute flux scale. However, the difference between the sky model and the true sky affects the gain solutions by absorbing this model incompleteness into the gain solutions. These gain errors, on all baselines, when applied to the data or the model lead to excess power on a wide range of spatial and frequency scales, also in the EoR window~\citep{10.1093/mnras/stw2277_patil2016,10.1093/mnras/stw1380_Barry_2016,10.1093/mnras/stx1221_Ewall,2018ApJ...868...63Ewall-wice}. Especially, very bright sources, such as Cas A and Cyg A, enter the FoV through their undulating sidelobes and could be a source of the excess variance. These bright sources have higher power than the fainter sources that enter the FoV through the rather slowly varying primary beam~\citep{2012ApJ...752..137M_Morales,2020MNRAS.493.1662M_Mertens_2020}. A single bright source is added coherently to the power spectrum, whereas many fainter sources (adding to the same total flux) is added incoherently. Thus, a few bright sources can dominate excess variance. Gain errors can both suppress structures that are not included in the model and add structures when applied to the model and subtracted from the data~\citep{Millad_2018}. 

\paragraph*{Diffuse emission: }
The unmodeled diffuse emission and partly polarized galactic emission can be a potential source of the excess variance as well. In our analysis, we tackle the problem in two ways: (1) applying $50\lambda$ cut on baselines to suppress the diffuse emission (the diffuse emission becomes problematic with baselines shorter than 100$\lambda$) and (2) applying GPR to remove spectrally smooth diffuse emission. The unpolarized part of the diffuse emission is spectrally smooth and can be removed well by GPR~\citep{10.1093/mnras/sty1207/gpr}. The polarization leakage of the diffuse to Stokes I, however, can introduce spectral fluctuations similar to the frequency structure of the 21 cm signal~\citep{Jelic_2015,10.1093/mnras/stv1107_Asad_2015,10.1093/mnras/stw1863_Asad_2016,Millad_2018}. The predicted level of leakage is estimated to be much smaller (i.e.,$\sim1\%$) than the observed level of excess variance~\citep{10.1093/mnras/stw1863_Asad_2016} and hence it should not be a main source of the excess variance at the current level. 

\paragraph*{Instrumental gain changes: }

Visibilities are affected by the instrument mainly due to their time-, frequency-, and direction-dependent gains variations.
DD-calibration corrects for these variations by solving gains independently for 122 directions as a function of frequency and time, thus incorporating the beam model and very slow ionospheric changes (see Section~\ref{par:ionosphere}).

Foreground sources that are far from the beam center and are not included in the sky model or not correctly subtracted, can leave chromatic PSF sidelobes in the central part of the beam being analyzed for the 21 cm signal, and also affect the gain solutions on longer baselines. These induced gain errors can affect the gain solutions on shorter baselines, similarly to ionospheric gain errors, which in our case are inferred from the longer baselines only ~\citep{10.1093/mnras/stw2277_patil2016,10.1093/mnras/stx1221_Ewall}.  The beam pattern changes more rapidly with increasing frequency. Therefore, the excess variance caused by the chromatic beam pattern can be indirectly investigated by their baseline and frequency dependence~\citep{10.1093/mnras/stw2277_patil2016}. 
LOFAR also has an instrumentally polarized response. If the calibration process does not take this into account, this effect can cause polarized signals to change the total intensity (i.e., the Stokes I component)~\citep{Jelic_2010,Jelic_2015}. This effect is known as "polarization leakage."\ The instrumentally polarized signal can introduce spectral fluctuations in power spectra, further exacerbated by Faraday rotation (see Section~\ref{par:ionosphere}),  and mimic or contaminate the 21 cm signal. This effect has been estimated to be very small for LOFAR-HBA and likely only contributes to $\sim$1\% of the observed excess variance~\citep{10.1093/mnras/stv1107_Asad_2015,10.1093/mnras/stw1863_Asad_2016,Jelic_2015}. In this work, we do not focus on analyzing the contribution of instrumental polarization.

\paragraph*{Gain regularization: }
During DD-calibration, we assume that direction-independent and non-smooth responses such as the structure in the band-pass response of stations and cable reflections~\citep{2013A&A...549A..11O,Beardsley_2016,Kern_2020,2020MNRAS.493.1662M_Mertens_2020} are taken out in the DI-calibration step and hence gain solutions should be spectrally smooth. The smoothness of gains is enforced by using \textsc{sagecal-co}~\citep{2016arXiv160509219Y_Yatawatta} with a third-order Bernstein polynomial~\citep{10.1093/mnras/stv596_Yatawatta_2015b}. 

Currently, gain solutions are only mildly regularized in the DI-calibration step due to the uncorrected band-pass and cable reflection structures in frequency. The remaining structure in the frequency direction can introduce small chromatic gain errors. These errors could result in spectral fluctuations in the EoR window and can be a source of excess variance. 

\cite{10.1093/mnras/stw2277_patil2016,10.1093/mnras/stw1380_Barry_2016,10.1093/mnras/stx1221_Ewall} showed that using small regularization constants (i.e., low regularization) can allow the gains to overfit the data. The extra frequency-dependent gain structure introduced by overfitting, when applied to the calibration model, can both leak power in the EoR window (excess variance) and lead to signal loss (bias). Previously,~\cite{Patil_2017} tried to tackle the latter problem by excluding the short baselines $<250\lambda$, used for 21-cm signal analyses, in DD-calibration. However, the gain-regularization parameters and the number of iterations used in~\cite{Patil_2017} were sub-optimal and resulted in high excess variance on the excluded baselines.~\cite{2020MNRAS.493.1662M_Mertens_2020} significantly improved the DD-calibration method by optimizing regularization values~\citep{suppression_paper_Mevius_2020}. The results showed that excess variance due to overfitting in the DD-calibration step is not a major contribution with the improved calibration scheme.  

\begin{figure}
        \includegraphics[width=\columnwidth]{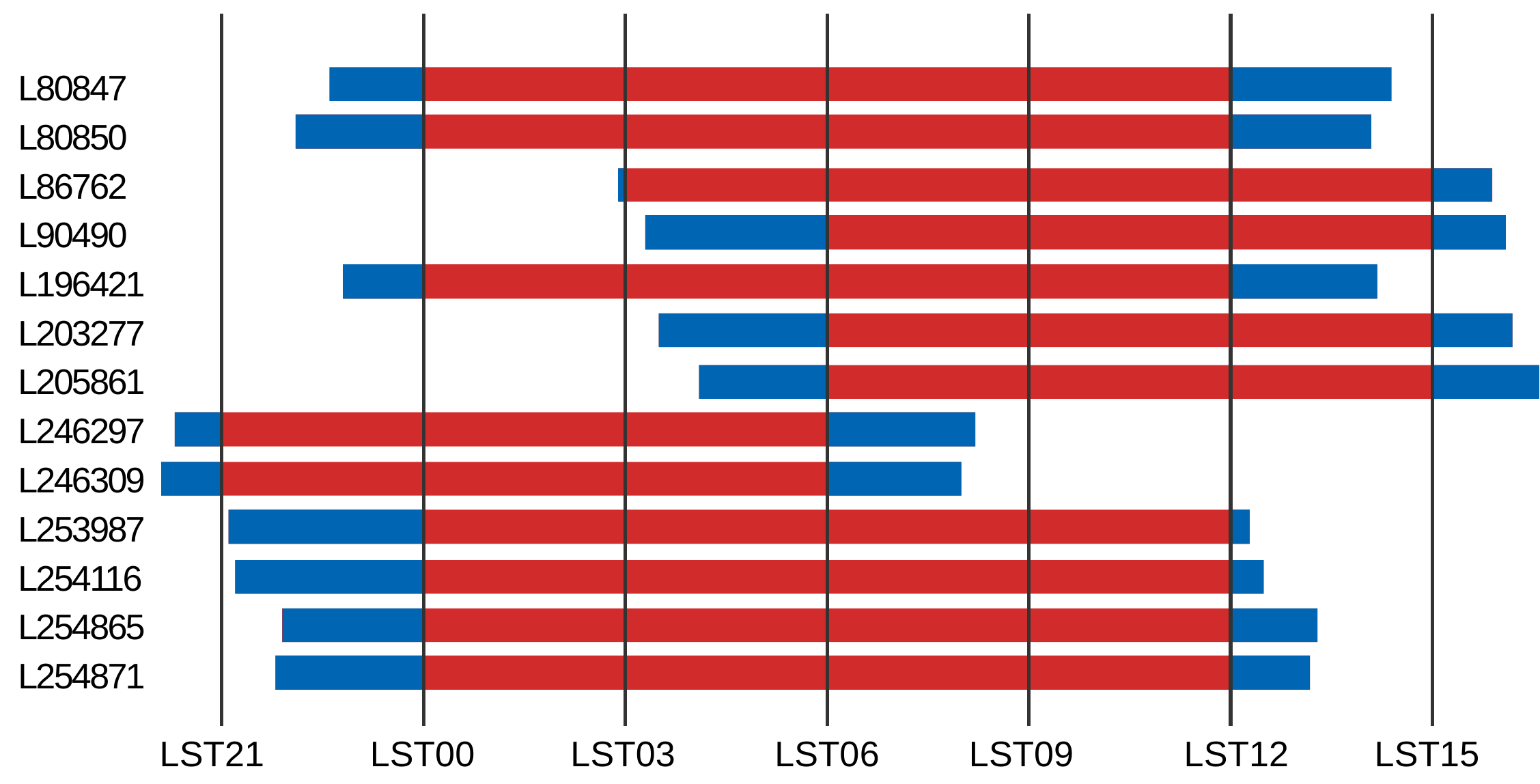}
    \caption{Duration of observations in LST for 13 nights. For analyzing the LST-dependence on excess variance, each observation yields three to four 3h LST slices (in red) and power spectra are estimated for each 3h slice, respectively. The data which do not fully cover a 3h LST range are discarded (in blue) and are not included in the analysis. }
    \label{fig:LST_slice}
\end{figure}

\subsection{Time and sky-model power dependence}

If excess variance depends on how much power the instrument receives, the expectation is for it to depend on local sidereal time (LST). To investigate this further, the data was binned in LST ranges and compared to sky-model power-spectrum simulations. 

\subsubsection{LST dependence} The sky-related effects on the 21 cm signal power spectra can be analyzed, for example, by analyzing the LST dependence of the excess variance. The power received by LOFAR is a product of the intensity coming from the sky and an averaged beam model which depends on receiver gains. Hence, an excess variance power that correlates with LST must come from a change in the product of the two (i.e., the sky intensity and the averaged beam). For similar LST ranges, the observed sky is similar (barring strong variable sources, which is a reasonable assumption at low frequencies). Therefore, for multiple observations within similar LST ranges, the difference between nights is likely to come from a difference in the instrument, $uv$-coverage, flagging, ionosphere, or RFI, and so on. However, effects that are uncorrelated between nights (e.g., RFI is more likely to correlate with UT, because it is caused by human activities) or are stochastic in nature (e.g., the ionosphere) are unlikely to show a strong LST dependence in the integrated power over multiple nights. 

In this work, we examine the correlation between the sky and the excess variance by looking at the variation of excess variance with LST. For this purpose, we sliced each observation into 3 hour (3h) LST ranges. The duration of the observations and the resulting 3 hour slices from each night are shown in Fig. \ref{fig:LST_slice}. We created a power spectrum for each slice after applying GPR foreground-removal with parameters optimized using data from the entire night to avoid overfitting and assuming that the power spectrum of the sky and signal does not change in a slightly rotated NCP. We estimated the excess variance level of each slice. By comparing individual LST slices, we were able to assess how the excess variance correlates with LST and also whether there are changes between nights within the same LST range. 

\subsubsection{Sky-model power simulations} Besides any LST dependence, to further assess whether the excess power correlates with received power, and from which sources, we simulated visibilities based on two sky models: (1) the two brightest radio sources in the northern sky, Cas A and Cyg A, which are outside the instrument's field of view ($\sim30^{\circ}$ and $\sim50^{\circ}$ away from the NCP, respectively); (2) the first 1416 bright sources that are used for DI-calibration and account for $\sim99\%$ of total sky model power together with A-team sources~\citep{de_Gasperin_2020_a-team} including Cas A and Cyg A. The simulations were created by matching the same observing times, sub-bands, and the $uv$-coverages of all 13 observations, with and without post DD-calibration flagging. The simulated visibility sets were also sliced into 3h LST bins to create power spectra. Finally, they were compared with power spectra from the actual observations. \\

\begin{figure}
    \centering
    \includegraphics[width=200px]{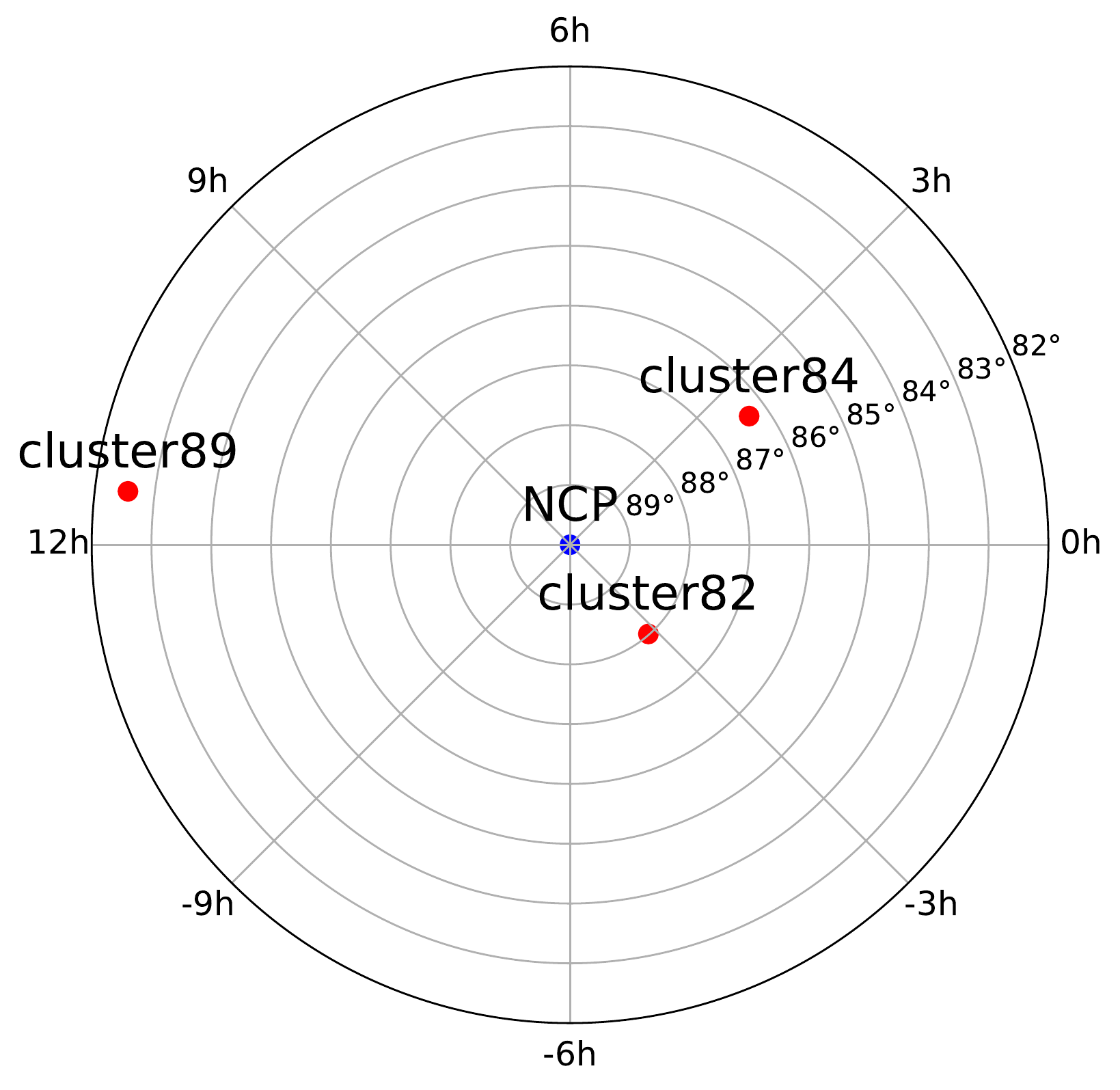}
    \caption{Distance of Cluster82, Cluster84, and Cluster89 from the NCP.}
    \label{fig:cluster_map}
\end{figure}


\section{Results}
\label{sec:results}

In this section, we present and discuss the correlations between the excess variance of the LOFAR-EoR power spectra and the metrics discussed in the Subsection~\ref{subsec:possible_causes}. We focus on gain calibration and sky-model and ionospheric related effects, as well as the impact of flagging.

\subsection{Gain solutions}
Before analyzing power spectra from LOFAR data, we first analyze gain solutions that are used to calibrate observed data. The gain solutions are obtained by \textsc{sagecal-co} as described in Section~\ref{sec:data_processing}. As discussed in Section~\ref{sec_sources}, gains are solved under the physics-motivated assumption that instrumental and ionospheric effects are smooth in frequency and on short time intervals. Therefore, irregular behaviors of gains, namely,\ non-smooth gains, over frequency or time, can be considered as gain errors. Such gain errors can propagate to power spectra of the 21 cm signal and contribute to the excess variance.

\subsubsection{Gain smoothness}
\label{subsubsec:gain_smoothness}

\begin{table*}
\caption{Information of three clusters used for DD-calibration gain analysis.}
\begin{threeparttable}
 \begin{tabular}{l l l r r r r l} 
 \hline
 \noalign{\smallskip}
 Cluster ID & RA & Dec & Number of & Solution interval & Distance from NCP & Number of & Size\\ 
  &  &  & solutions & [min] & [degree] & sources & [degree]\\ 
 \hline
 \noalign{\smallskip}
 Cluster82 & $-03^{\text{h}} 14^{\text{m}} 46.2008^{\text{s*}}$ & $88\degr 00\arcmin 51.1354\farcs$ & 4 & 5 & 2.0 & 438 & 1.21 \\ 
 Cluster84 & $02^{\text{h}} 22^{\text{m}} 44.4171^{\text{s}}$ & $86\degr 18\arcmin 43. 8404\farcs$ & 8 & 2.5 & 3.7 & 105 & 0.03$^\text{**}$\\
 Cluster89 & $11^{\text{h}} 32^{\text{m}} 26.8737^{\text{s}}$ & $82\degr 33\arcmin 10.4620\farcs$ & 1 & 20 & 7.4 & 370 & 2.63\\ [.8ex] 
 \hline
 \noalign{\smallskip}
\end{tabular}
  \begin{tablenotes}
    \item[*] RA ranges from $-12^{\text{h}}$ to $+12^{\text{h}}$.
    \item[**] Cluster84 is a bright source, 3C61.1.
  \end{tablenotes}
\end{threeparttable}
\label{table:clusters}
\end{table*}

\begin{figure*}
    \includegraphics[width=\textwidth]{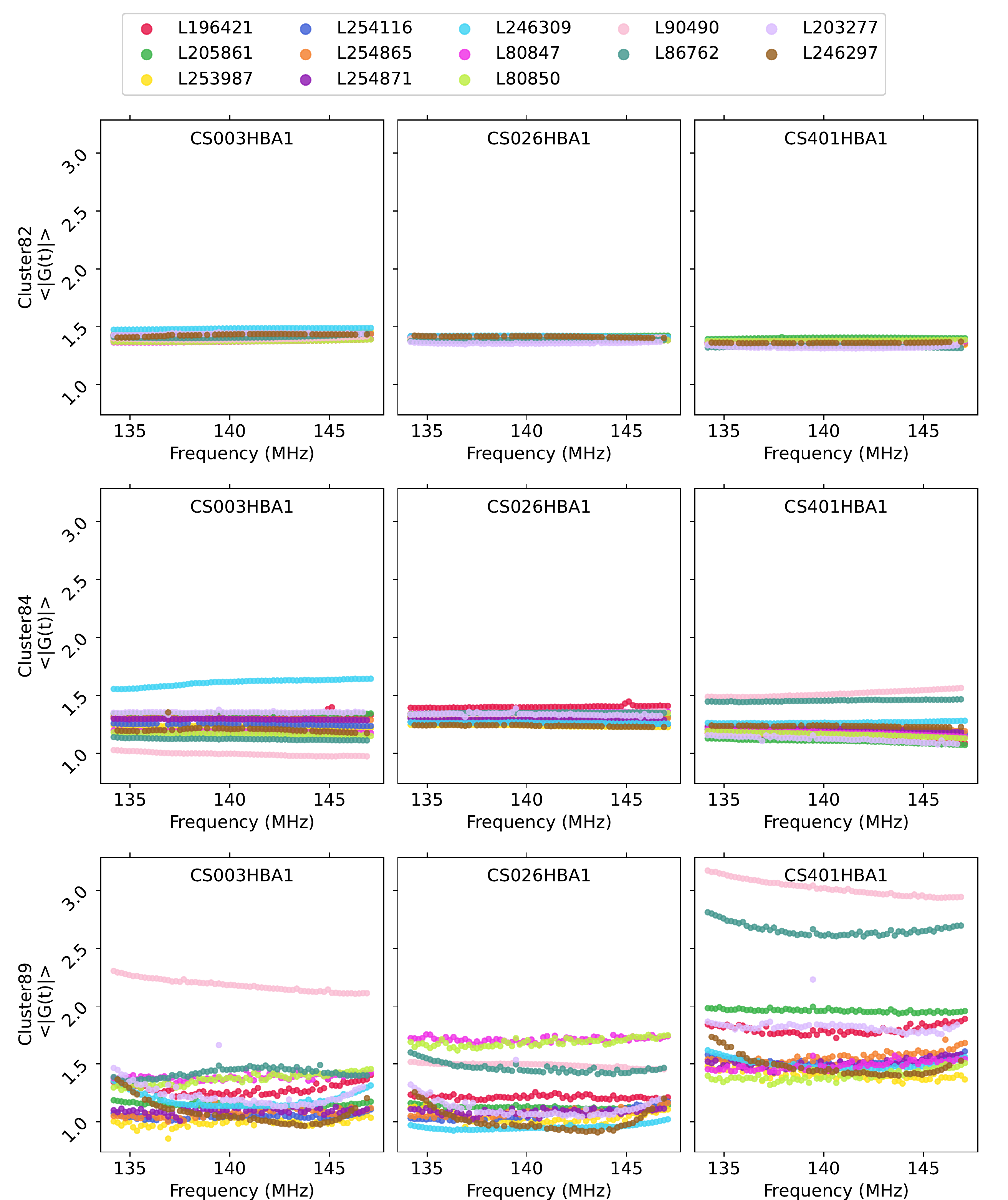}
    \caption{Time-averaged DD-gains as a function of frequency from 13 night LOFAR observations (in different colors) in Cluster82 (top row), Cluster84 (middle row), and Cluster89 (bottom row). The calculated time averages are separately plotted for three chosen core stations --- CS003HBA1 (left column), CS026HBA1 (middle column), and CS401HBA1 (right column), respectively. The color code denotes different nights and each data point indicates a time average at a fixed frequency and a night. Time-averaged gains are smoother over frequency as the cluster is closer to the NCP (from bottom to top).}
    \label{fig:t_avg}
\end{figure*}

\begin{figure*}
    \includegraphics[width=\textwidth]{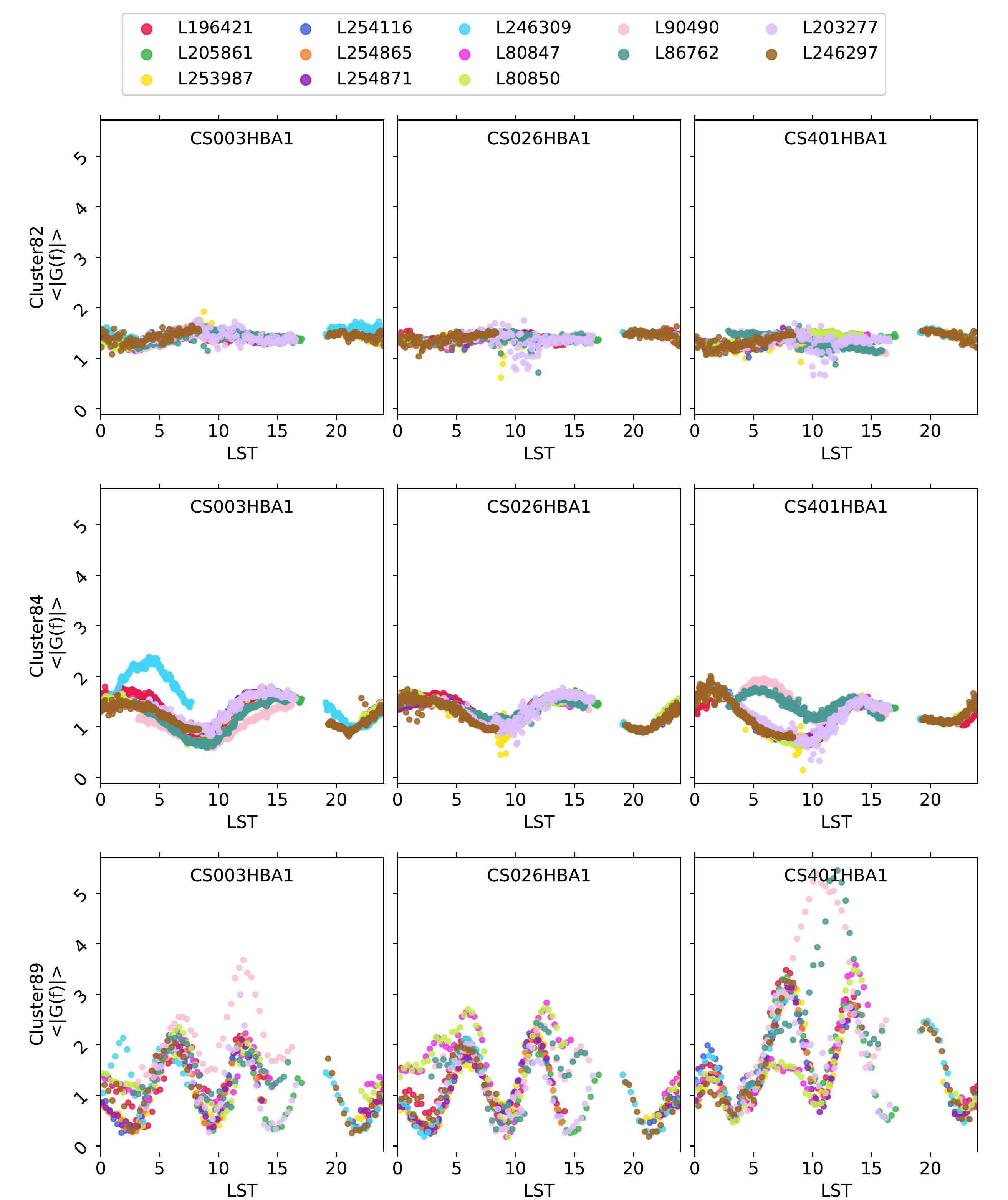}
    \caption{Frequency-averaged DD-gains as a function of time from 13 night LOFAR observations (in different colors) in Cluster82 (top row), Cluster84 (middle row), and Cluster89 (bottom row). The calculated time averages are separately plotted for three chosen core stations --- CS003HBA1 (left column), CS026HBA1 (middle column), and CS401HBA1 (right column), respectively. The color code denotes different nights and each data point indicates a frequency average at a fixed time and a night. Frequency-averaged gains fluctuate more rapidly over LST as the cluster is farther away from the NCP (from top to bottom).}
    \label{fig:f_avg}
\end{figure*}

We calculated and analyzed averages of gain solutions over time and frequency, respectively, to assess their smoothness level and investigate whether there are possible errors in gains. We limited ourselves to three out of the 122 directions: Cluster82, Cluster84, and Cluster89 (see Fig.~\ref{fig:cluster_map}), which are inside, around, and outside the full-width-half-maximum (FWHM) of LOFAR-HBA tiles being ~$\sim3.8^{\circ}$ at 150\,MHz~\citep{van_Haarlem_2013_refId0}, respectively. Detailed information about the three clusters is summarized in Table.~\ref{table:clusters}. Figures~\ref{fig:t_avg} and \ref{fig:f_avg} show the DD-gain average over time and frequency of 13 observations for the selected three clusters and three core stations, respectively. The three stations are selected based on their location: CS003 resides on a 320 m diameter island at the core of LOFAR, known as the “Superterp;” CS026 and CS401 reside on two opposite sides (in the east and west) of CS003 outside the Superterp within $\sim1$ km diameter from the LOFAR core. The DD-gain solutions calibrate measured visibilities for each station, polarization, time, and frequency. To obtain the gain average over time or frequency, each polarization component (XX, XY, YX, YY) is added in quadrature to obtain the total intensity. Afterwards, gains are averaged over time or frequency for each station, respectively. At this stage, the averaged gains have two dimensions, namely, (station, time) or (station, frequency). We note that these gains are not applied to the sky-model and subtracted from the data, but those from the Bernstein-polynomial fit. The latter is sufficiently smooth not to cause the observed excess variance \citep{2020MNRAS.493.1662M_Mertens_2020}; hence, any gain averages and variances only act as an indicator of other gain effects that might affect the power spectrum.

\paragraph*{Gain smoothness over frequency:}
Time-averaged gains do not show a strong frequency dependence for the three selected source clusters (i.e.,\ the direction). For specific observations and stations, for example, the time-averaged gains for observation L246309 in Cluster84 and CS003HBA1 (the cyan line on the left column and middle row in Fig.~\ref{fig:t_avg}) increases as frequency increases, but this tendency only appears in a few cases. It is rather random and there is no strong correlation between gains and different clusters or nights. The time-averaged gains are in general smooth between neighboring frequency intervals and we do not see huge jumps in the gain average for a selected cluster, station and night. 
We notice that the time-averaged gain in Cluster89 (bottom row in Fig.~\ref{fig:t_avg}) has less smooth curves compared to ones from Cluster82 and Cluster84 (first two rows in Fig.~\ref{fig:t_avg}). A cluster located farther from the NCP have gains solutions that are less smooth over frequency compared to gains for a cluster closer to the NCP. We see that in a given cluster and a station, gains per night show a slight increase towards its two frequency ends and this increase is more obvious in outer clusters (from top to bottom in Fig.~\ref{fig:t_avg}), where the S/N is lower~\citep{2020MNRAS.493.1662M_Mertens_2020,suppression_paper_Mevius_2020}. 
The different level of smoothness in gains is possibly also related to the different solution time interval in each cluster. Cluster89 has one solution in every 20 min while Cluster82 has 4 solutions and Cluster84 has 8 solutions per 20 min, respectively. This means that a single data point in Cluster82, Cluster84 and Cluster89 is a time-averaged gain over 5 min, 2.5 min, and 20 min, respectively. Finer time intervals are better at catching small variations in ionospheric and beam effects. 

\paragraph*{Gain smoothness over LST:} 
Figure~\ref{fig:f_avg} shows frequency-averaged gains as a function of LST per night and station. The frequency-averaged gains are calculated in a way similar to how the time-averaged gain is calculated, however, this time we average over frequency in the final step, instead of time. There are gaps among data points appearing around LST15-20 in all three clusters because the starting times of some observations are later in LST and observations end earlier in LST. We note that gains are more sparsely distributed in Cluster89 compared to ones in Cluster82 and Cluster84 due to the longer solution time interval. In a fixed cluster, the station-to-station difference (between columns in Fig.~\ref{fig:f_avg}) is rather small compared to the cluster difference (between rows in Fig.~\ref{fig:f_avg}).
For a fixed station and a night, the gain averaged over frequency per night shows a clear LST-dependence. The frequency-averaged gain typically shows a sinusoidal behavior with LST. This is because the beam shape changes over time due to the Earth's rotation. The smoothness of the sinusoidal gain curves depends largely on the cluster (i.e., the direction, between rows in Fig.~\ref{fig:f_avg}). Cluster84 is $\sim3.7^{\circ}$ away from the NCP and the FWHM of HBA tiles in core stations is $\sim3.8^{\circ}$ at 150 MHz~\citep{van_Haarlem_2013_refId0}. As a result, Cluster84 (middle row in Fig.~\ref{fig:f_avg}) is still within the FWHM of HBA tiles and the corresponding gains do not change rapidly compared to Cluster89 (bottom row in Fig.~\ref{fig:f_avg}), which is farther away from the NCP than Cluster84. We observe that as the cluster is located farther away from the NCP, the frequency-averaged gain fluctuates more rapidly (from top to bottom row in Fig.~\ref{fig:f_avg}) and vice versa.

\paragraph*{Night to night correlation:} 
The LST dependence of the gains is very strong and in a fixed cluster, the sinusoidal behavior of gains is similar between stations as well except for some specific nights appearing as outliers. We note that some specific nights keep appearing as outliers in the same station of multiple clusters. For instance, two observations L90490 and L86762 (in pink and teal, respectively) constantly deviate from other nights in station CS401HBA1 for gain averages over time and frequency for Cluster84 and Cluster89 (last column on the second and third panels in Fig.~\ref{fig:t_avg} and~\ref{fig:f_avg}). Similarly, L80847 and L80850 in Cluster89 and CS026HBA1 appear as outliers. We suspect that the "outlier nights" come from some station-based effects, such as station orientations or broken tiles in stations that occurred for specific observations. To check this point, we compare tile flagging of core stations between observations and figure out that observations can be categorized into three groups based on their similarity of tile flagging: (1) L80847 and L80850; (2) L86762 and L90490; and (3) the rest. 
We note that observations in the same group were taken closely in UTC (in Table.~\ref{table:obsv}) and they tend to have more similar tile flagging. As we expected, the outliers exactly come from the tile flagging groups (1) and (2) which have different tile flagging compared to other observations. This supports that the night to night difference in gain average is mainly due to the tile flagging difference between observations, which leads to differences in their station beams and, hence, their DD gains.

\begin{figure*}
    \centering
    \includegraphics[width=450px]{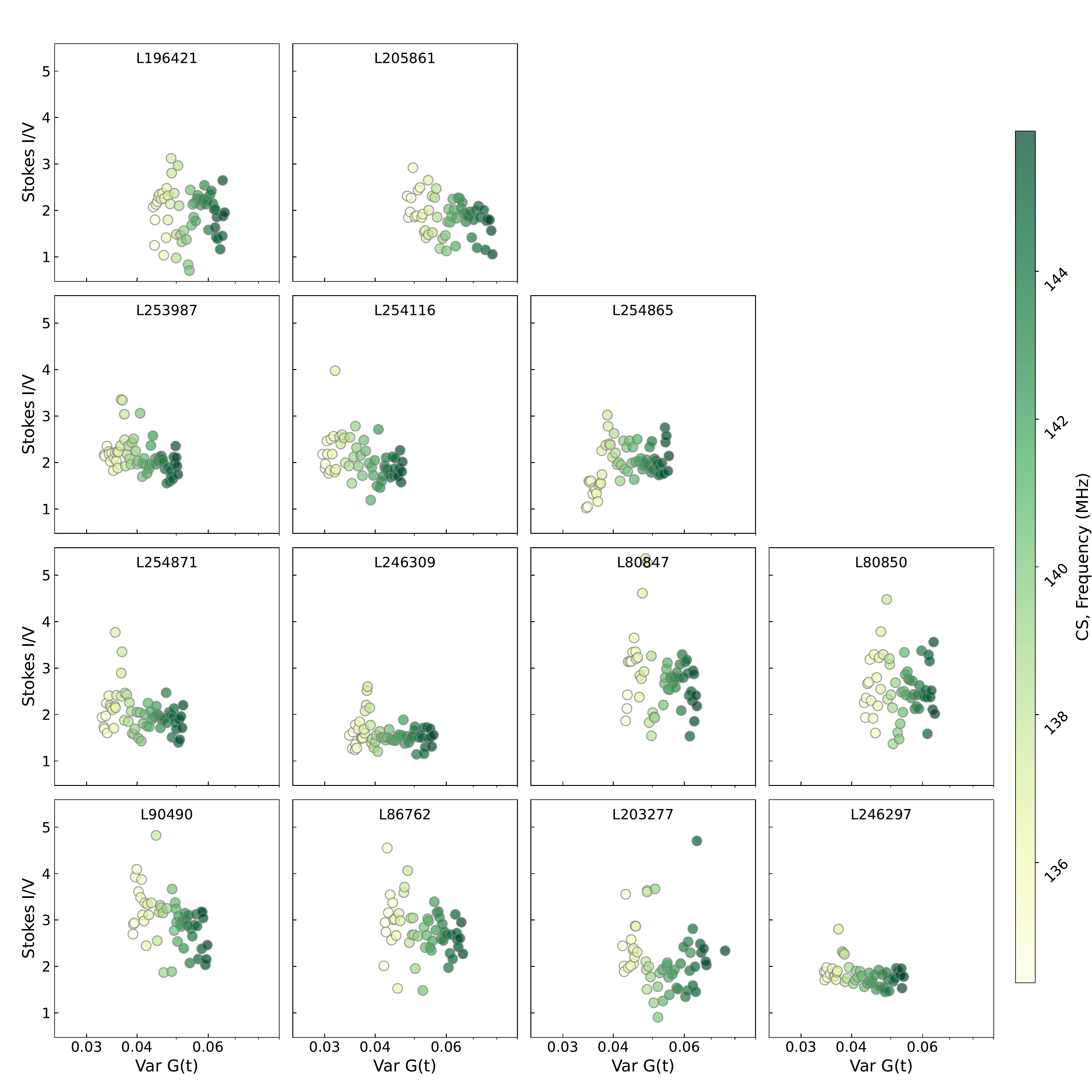}
    \caption{Excess variance as a function of DD-gain variance over time in Cluster84 for 13 nights. The color denotes frequency. Each data point in a fixed night is a gain variance over time at a fixed frequency with the corresponding Stokes I/V ratio. There is no strong correlation between the excess variance and gain variance over time. However, the gain variance has a frequency dependence: higher frequencies tend to have higher gain variances.}
    \label{fig:excess_var(g)_C84}
\end{figure*}
\subsubsection{Correlation between gain solutions and excess variance}

The instrument and the sky should have spectrally smooth gains; hence, any abrupt jump in the gain average or variance over frequency or time is an indication of excess noise\footnote{We applied the Bernstein-polynomial gain fits to the sky model, before subtracting it from the data. \citet[][]{2020MNRAS.493.1662M_Mertens_2020} have shown these lead to spectrally smooth model power spectra; hence, any non-smooth spectral gains must be due to effects other than the instrument, possibly numerical or, still, a reflection of an incomplete sky model partly absorbed by the gain. Therefore, non-smooth gains are not the cause of the excess noise, but an indicator that some effects are not accounted for in the data model.}. To assess whether there is a correlation between excess variance and gain errors, we plotted excess variance against the DD-gain variance.
Four polarization components in gains, solutions were added in quadrature to calculate a "total intensity" equivalent. The gains were then averaged over core stations to increase the S/N and the variance is calculated along the time axis; hence, the calculated variance is given as a function of frequency. 

\begin{figure*}
    \centering
    \includegraphics[width=400px]{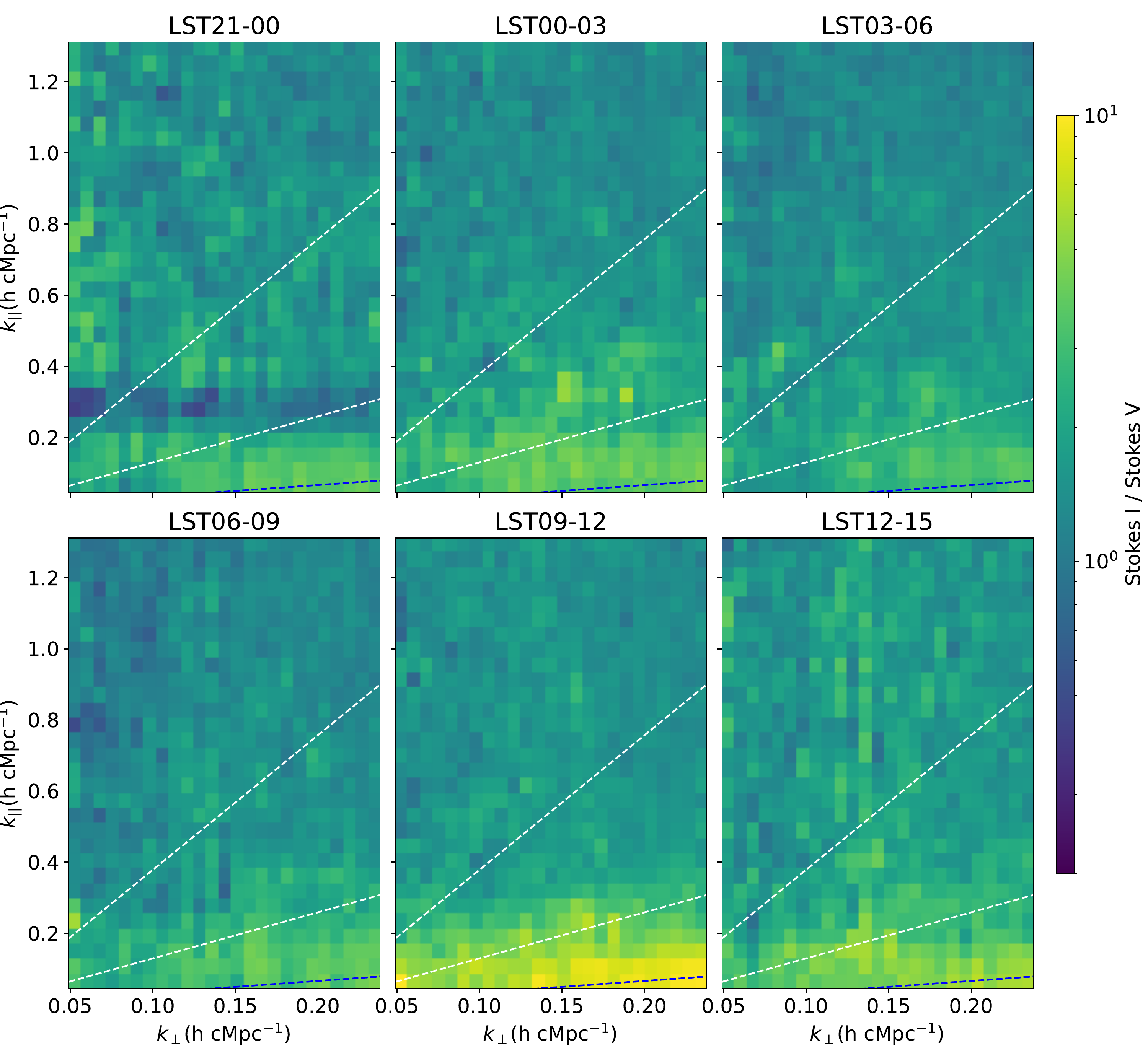}
    \caption{Median value over 13 nights of the ratio (i.e.,\ Stokes I over Stokes V) of the cylindrical power spectra per LST bin. We define this as a cylindrical "excess variance" power spectrum. The dashed lines indicate the 5\degr~(the primary beam; bottom), 20\degr~(middle),~and instrumental (top) horizon delay lines, respectively. The excess variance is especially high ($>5$) in the wedge region below the $\sim$20\degr~delay line in all LST bins. The excess variance has an LST dependence and LST\,09-12 has the highest excess power in the wedge region compared to other LST bins.}
    \label{fig:13night_2dPS}
\end{figure*}
To estimate the excess variance in the 21 cm power spectra, we used the Stokes I and V components of the spatial power spectra. The spatial power spectra were obtained by performing Fourier transform along the spatial coordinate ($u,v$) on measured visibility cubes. The spatial power spectrum is given as a function of frequency and baseline ($k_\perp$). We averaged the spatial power spectrum over baseline for the Stokes I and V, respectively. For a given frequency, we took the ratio of the averaged Stokes I and V. Finally, the gain variance over time and the Stokes I/V ratio was mapped by matching frequency. We note that a disadvantage of this analysis is that the baseline dependence on gains and excess variance will be erased because we are averaging over baselines to calculate the Stokes I/V ratio as a function of frequency. In a forthcoming paper, we will examine excess noise in more detail as a function of baseline and delay.

Figure~\ref{fig:excess_var(g)_C84} shows excess variance as a function of DD-gain variance over time in Cluster84 per night, respectively. We immediately note that the excess variance does not strongly depend on either frequency or gain variance. Rather, the gain variance over time increases as frequency increases and this is consistent for all 13 night observations. Since the primary-beam FWHM is inversely proportional to frequency and gets narrower as frequency increases, we can expect that gains fluctuate more for higher frequencies, and the gain variance increases as well. This is indeed seen in the figure. 

However, our current analysis is limited to a relatively narrow frequency range (134.1-147.1 MHz) compared to the three-fold broader frequency range (121.8-159.3 MHz) that was previously presented in~\cite{Patil_2017}. To be able to investigate the frequency dependence of gain variance, we need to add more data with a wider frequency range. In the future, we will extend our analysis to a wider redshift range. This extended analysis can give a more clear correlation between excess variance and frequency. However, the current conclusion is that excess variance does not correlate with gain variance. If anything, it has a slight anti-correlation for some nights.

\subsection{Sky-related effects}
\label{subsec:sky_effects}
In this subsection, the impact of sky-related effects on excess variance is investigated by performing jackknife tests on the observed 21 cm power spectra and the simulated power spectra. Individual observations and simulated visibility sets are divided into 3h LST bins (see Fig.~\ref{fig:LST_slice}) and power spectra are created for each of them. By limiting observations into 3h LST bins, we expect to be able to obtain equivalent PSFs and primary beams for different observations per 3h LST bin. This way, we can focus on investigating the impact of some bright sources which might be more prominent at a given LST.

\subsubsection{LST dependence}

Each night was divided into 3h LST bins and the data that do not fully cover the full 3h LST were discarded. Based on these 3h LST bins, we created visibility cubes and estimate their power spectra. We note that the number of observations in each 3h LST bin is not equal for each night (Fig.~\ref{fig:LST_slice}).
We also created power spectra for 1h and 6h LST bins, respectively, but find that the 3h bins have the optimal duration. The 1h LST bins have very limited $uv$-coverage and the generated power spectra end up containing much higher excess variance and residual foregrounds compared to power spectra from longer observations. The 6h LST bins have a good $uv$-coverage but result in only 1 or 2 bins per night; hence, the 6h LST bins do not provide sufficient information about LST dependence. The 3h LST bins still have a relatively good $uv$-coverage and provide us with enough information about LST dependence. 

Figure~\ref{fig:13night_2dPS} shows the median of 13 nights of cylindrical excess variance power spectra (i.e., the residual Stokes I and V ratio) per LST bin. We see that even after the DI-, DD-calibration, and foreground removal with GPR, there is still significant excess variance in power spectra, especially in the foreground wedge region (below the $\sim$20\degr~delay lines). The excess variance can reach up to $\sim$10 times the expected Stokes-V power (thermal noise) in the primary beam region, namely, the area below the $\sim$5\degr~delay lines in LST09-12. In LST21-00, we see a horizontal stripe ($k_{||}\sim0.3 \text{ }hc\text{Mpc}^{-1}$) of lower excess variance. In this particular LST bin, we have fewer samples compared to other LST bins, namely, only two observations (L246309 and L246297). The power in Stokes V is higher in L246309 around $k_{||}\sim0.3 \text{ }hc\text{Mpc}^{-1}$ and when we take the median of the Stokes I/V ratio, this higher power in Stokes V from one night stands out. 
This is not limited in LST21-00 of L246309 because we also see a similar stripe around $k_{||}\sim0.3 \text{ }hc\text{Mpc}^{-1}$ in the full night excess variance power spectrum of L246309 in Fig.~\ref{fig:full_night_excess}. It is not clear why the power of Stokes V in L246309 is higher than other nights. Adding more samples possibly removes the stripe and gives a cleaner EoR window (the region above the $\sim90^\circ$~delay line).
\begin{figure}
    \centering
        \includegraphics[width=230px]{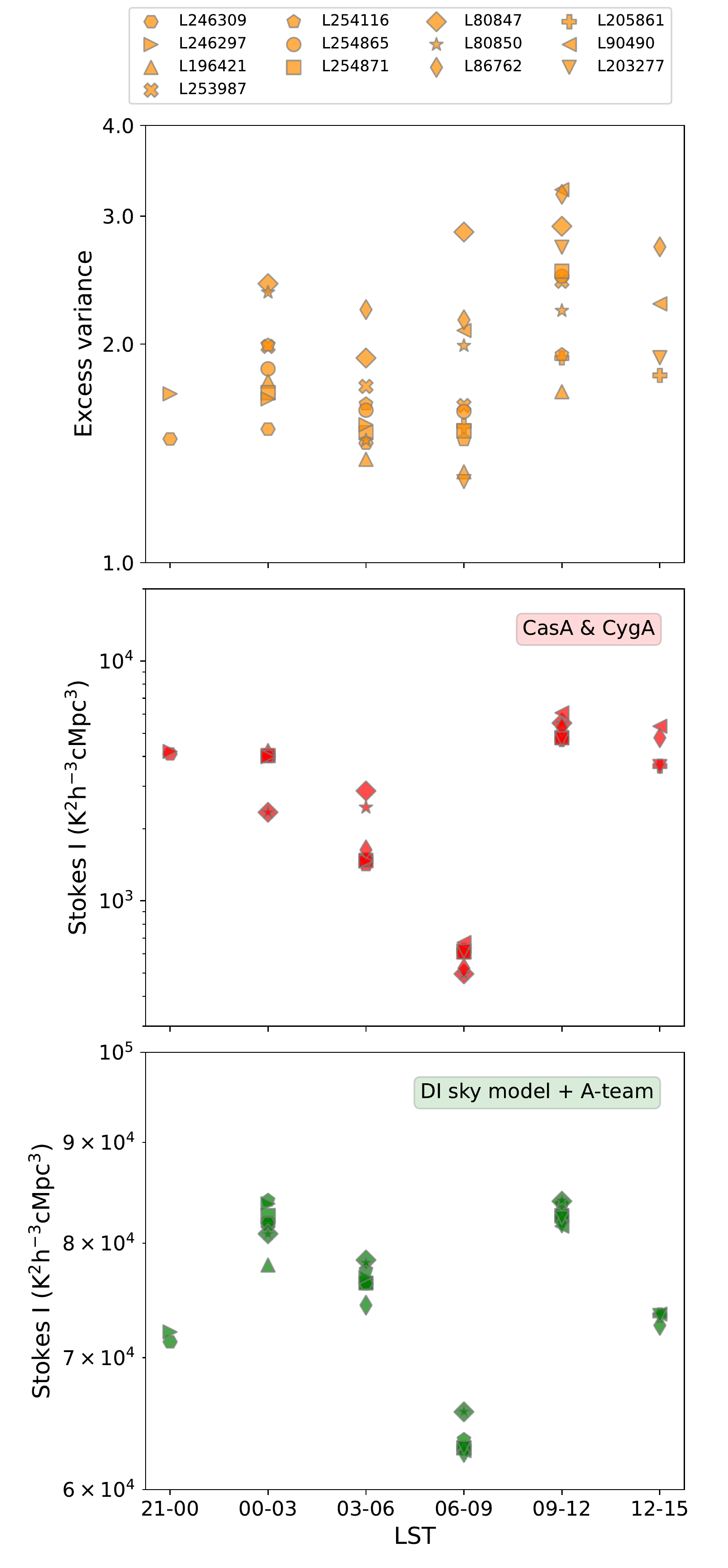}
    \caption{LST-dependence of excess variance and two sky models. {\bf Top:} Excess variance averaged over $k$ per night and per LST in orange. {\bf Middle:} Simulated Stokes I cylindrical power spectra averaged over $k$ per night and LST with Cas A and Cyg A in red. {\bf Bottom:} Simulated Stokes I cylindrical power spectra averaged over $k$ per night and LST with the DI-calibration model (including the 1416 brightest components that account for $\sim99\%$ of the flux from our sky model) and A-team sources in green. Different markers indicate different observations. The difference between nights in simulations mainly comes from the different number of sub-bands per night (see Table.~\ref{table:obsv}) and flagging based on $uv$-cells. The excess variance from observations and the simulated Stokes I power from the two models (in middle and bottom) shows a similar LST-dependence.}
    \label{fig:excess_vs_LST}
\end{figure}
The excess variance in the wedge below the $\sim$20\degr~delay lines changes drastically, depending on LST. LST09-12 shows the highest excess variance level in the wedge region compared to other LST bins. LST03-06 and LST06-09, on the other hand, show relatively a lower excess variance level. 
To further quantify the LST-dependence of the excess variance, we calculate the excess variance by averaging the median ratio of the cylindrical residual Stokes I and V over $k_\perp$ and $k_{||}$. The top panel in Fig.~\ref{fig:excess_vs_LST} shows the excess variance as a function of LST for all 13 nights. The excess variance varies with LST and it has the lowest values around LST03-06 and LST06-09. The excess variance shows a slight increase for other LST bins. This suggests that the excess variance depends on LST and is correlated with the sky. Both RFI and ionosphere are less likely to correlate with LST, that is, the PSF and primary beam. Different LST bins have different PSFs and primary beams and hence this difference is more prominent for bright sources in the sky seen by the instrument. RFI and ionosphere do not suffer from changing PSFs and primary beams. The scatter within each LST bin, however, is considerable and this is further investigated below. \vfill

\subsubsection{Simulation}
\label{subsub:simulation}
\begin{figure*}
    \includegraphics[width=250px]{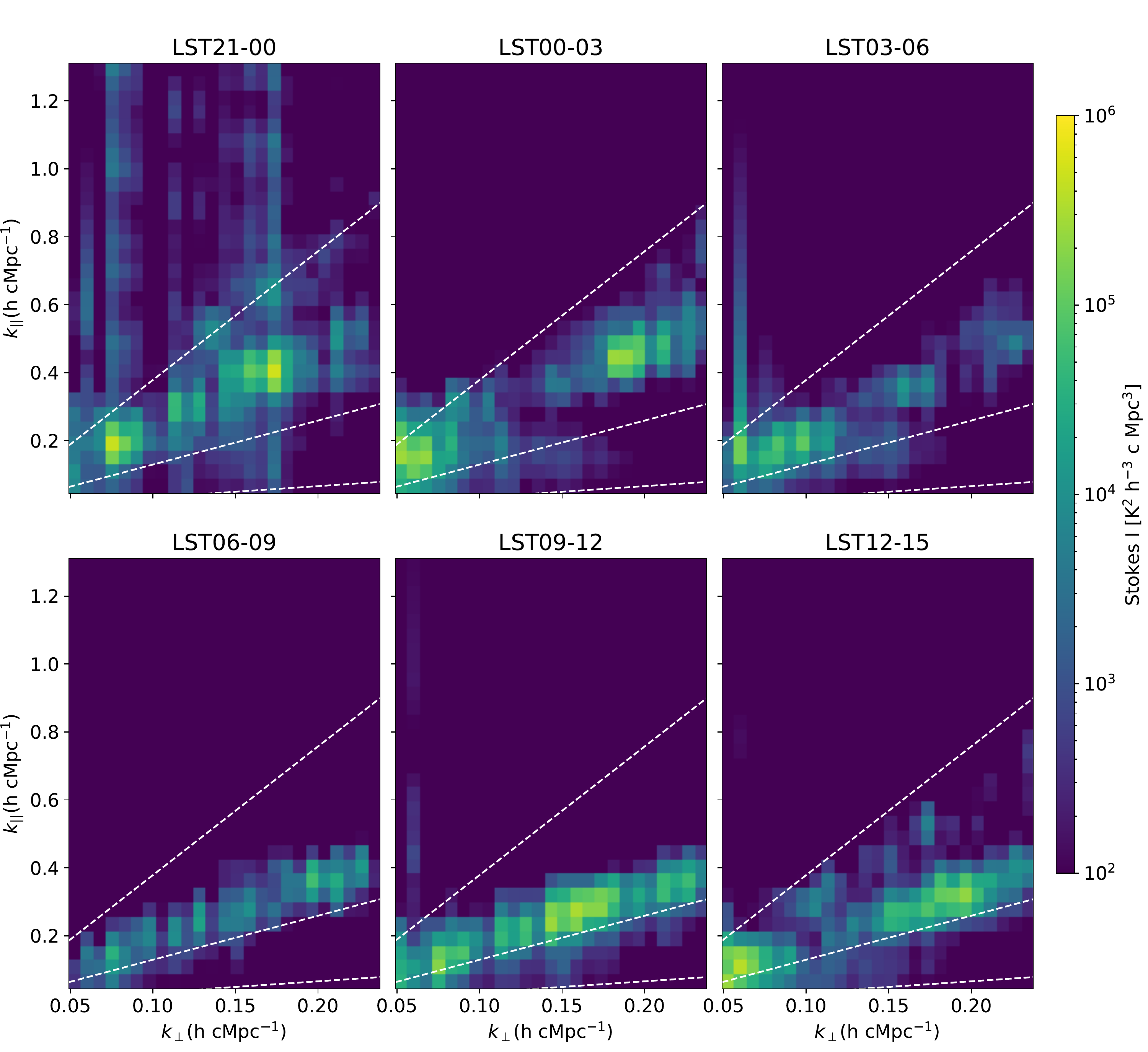}
    \includegraphics[width=250px]{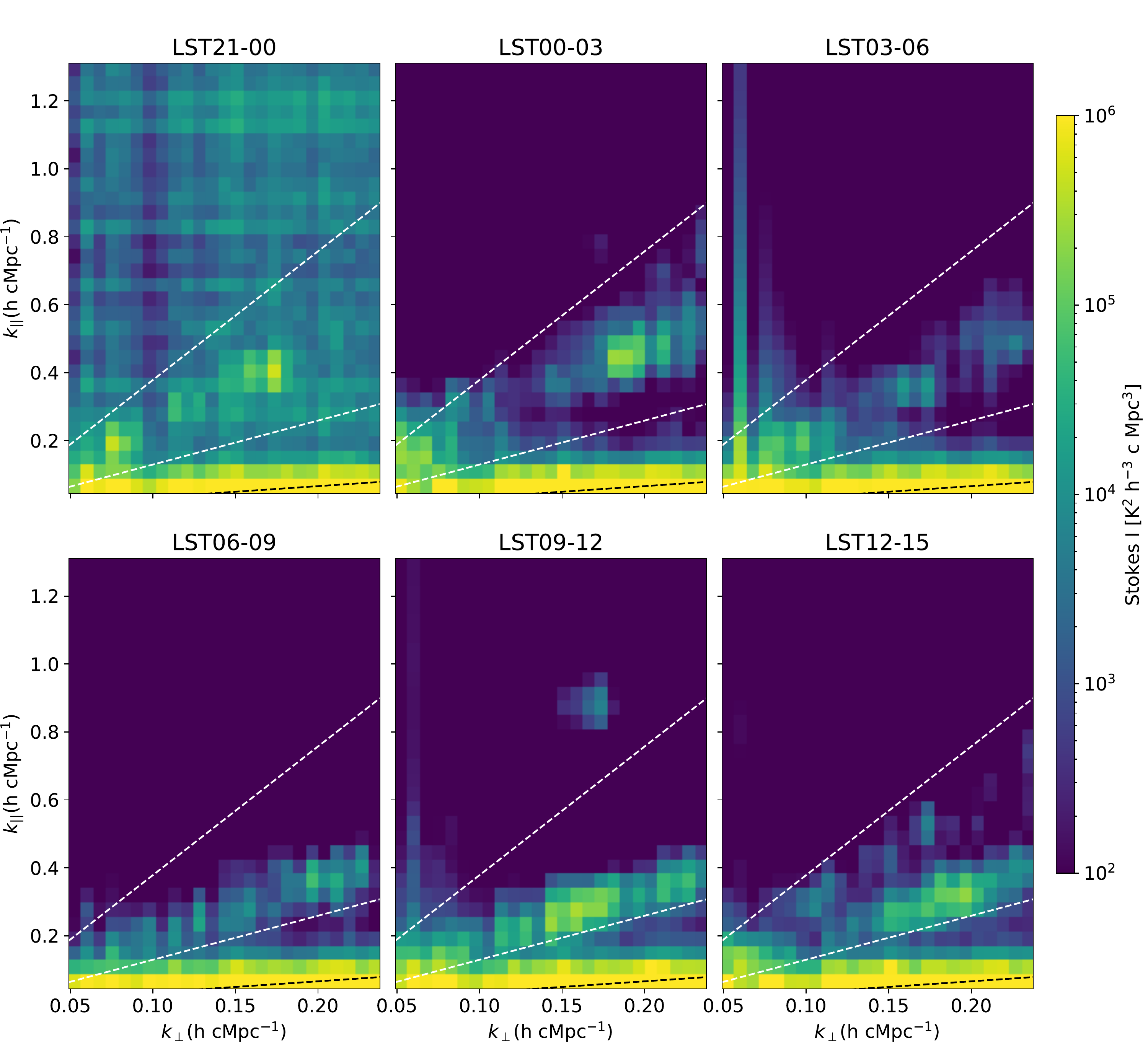}
    \caption{Median cylindrical Stokes I power spectra of the 13 night simulations with two different sky models per LST bin based on the 13 observations including $uv$-cell flagging. {\bf Left:} Cas A and Cyg A model. Dashed lines indicate, from bottom to top, $5^\circ$~(the primary beam), $20^\circ$~and instrumental horizon delay lines. {\bf Right:} the DI-calibration model (including the 1416 brightest components that account for $\sim99\%$ of the flux from our sky model) and A-team sources. LST03-06 and LST06-09 in both simulations show lower power compared to other LST bins and this trend is consistent with what we have seen with the excess variance in~Fig.\ref{fig:13night_2dPS} and the top panel of~Fig.\ref{fig:excess_vs_LST}.}
    \label{fig:simulated_CasA+CygA_PS}
\end{figure*}   
To verify whether the LST dependence of the excess variance is correlated with the power spectrum of the sky model, we simulated the visibilities and performed a similar test on the simulations. Our current sky model has more than 28,000 components (see Sec.~\ref{sec:data_processing}). However, the dominant contribution comes from very bright sources such as Cas\,A and Cyg\,A, therefore, we did not use the entire sky model in the simulation but, instead, we limited the simulation to the use of (1) Cas A and Cyg A only; (2) the 1416 brightest components used for DI-calibration and A-team sources, which contribute about 99\% to the sky power spectrum. We used \textsc{dppp}\footnote{\url{https://www.astron.nl/lofarwiki/doku.php?id=public:user_software:documentation:ndppp}} to predict visibilities including the LOFAR-HBA beam model. In this step, we used the \textsc{unflag} option to remove all the antenna flagging from the observation, mainly from RFI flagging, as simulated data do not have RFI issues (see Appendix~\ref{appendix2} for more discussion). The predicted visibilities were then sliced in LST as we sliced the observations and then they were subsequently imaged by \textsc{wsclean} using baselines longer than $40{\lambda}$ and shorter than $260{\lambda}$. Finally, the created image cubes were used to estimate the cylindrical power spectra in Stokes I, at this stage, we also perform post-processing flagging to filter out some $uv$-cells with bad $uv$-coverage. This extra step is needed because we are now working on relatively short observations (3 hours long) and the $uv$-cells with bad coverage can add extra power into the EoR window. The impact and necessity of flagging on simulated data (RFI unflagging and $uv$-cell flagging) are discussed in Appendix~\ref{appendix2}.
Figure~\ref{fig:simulated_CasA+CygA_PS} shows the cylindrical power spectra of the Stokes I from two different simulations with sky models (1) and (2) mentioned above. The left panel shows that the power from Cas A and Cyg A is concentrated under the instrumental horizon delay lines (top dashed lines) and above the primary beam (bottom dashed lines). Adding more sources to simulation, in the right panel, adds more power into the region below the horizon, especially to the primary beam region (below the bottom dashed lines). 

\paragraph*{Simulation cross-checks between nights: }
the $k$-averaged simulated Stokes I power spectra with the two models per LST and per night are shown in Fig.~\ref{fig:excess_vs_LST} (middle and bottom). The typical power in Stokes I increases about an order of magnitude from the Cas A and Cyg A only model to the DI-calibration model with A-team sources. We notice that even within a fixed LST bin, the simulated power can be different from night to night. This night-to-night difference happens because our simulations are based on real observations. We follow the same observing times and beam patterns (including antenna flagging, sub-bands) of real observations. This implies that (1) antennas in stations are flagged differently for each night; (2) the 3h LST bins based on the observing time are not exact and can have up to 10-sec errors, because of the 10 sec time interval in data and the difference in start and end times between observations; (3) the number of sub-bands differs from night to night. While the difference in LST bins and the number of sub-bands per night are already defined, the station flagging difference is more due to defects in the instrument at a specific time in a real observation that do not exist in a simulation. To make our simulations more realistic, we manually "unflag" antennas from RFI flagging during the simulation and then flag $uv$-cells with bad coverage at the stage of power spectrum estimation as discussed earlier. In this way, we do not flag extra stations in simulations. Typically, up to 5\% of total visibilities are filtered out in the $uv$-cell flagging. The 10 sec of difference in 3h LST bins should not lead to a significant difference in power because the modeled power is smooth over time and 10 seconds accounts for less than 0.1\% of the 3 hour observing run. Therefore, we conclude that the difference between nights mostly originates from the difference in the number of sub-bands and from the $uv$-cell flagging. We also notice that there might be differences in tile and antenna flagging between nights, which could also contribute to the difference between nights per 3h LST data, however, single tile or antenna flagging in simulations cannot be switched off easily at the current level; hence, we do not discuss this case in this paper. 

\paragraph*{Correlation between simulations and the excess variance: }

The $k$-averaged power spectra of the excess variance and the simulated Stokes I from two models are compared (from top to bottom in Fig.~\ref{fig:excess_vs_LST}). We note meaningful similarities in the progression of the averaged power spectra over LST between the excess variance and the simulations. The LST-dependence we first see from the excess variance in Fig.~\ref{fig:13night_2dPS} and Fig.~\ref{fig:excess_vs_LST} (top) also appears in the simulations in Fig.~\ref{fig:excess_vs_LST} (middle and bottom) and Fig.~\ref{fig:simulated_CasA+CygA_PS} with the minimum power appearing at LST06-09, especially when comparing their lower envelopes. 

\begin{figure}
    \centering
        \includegraphics[width=240px]{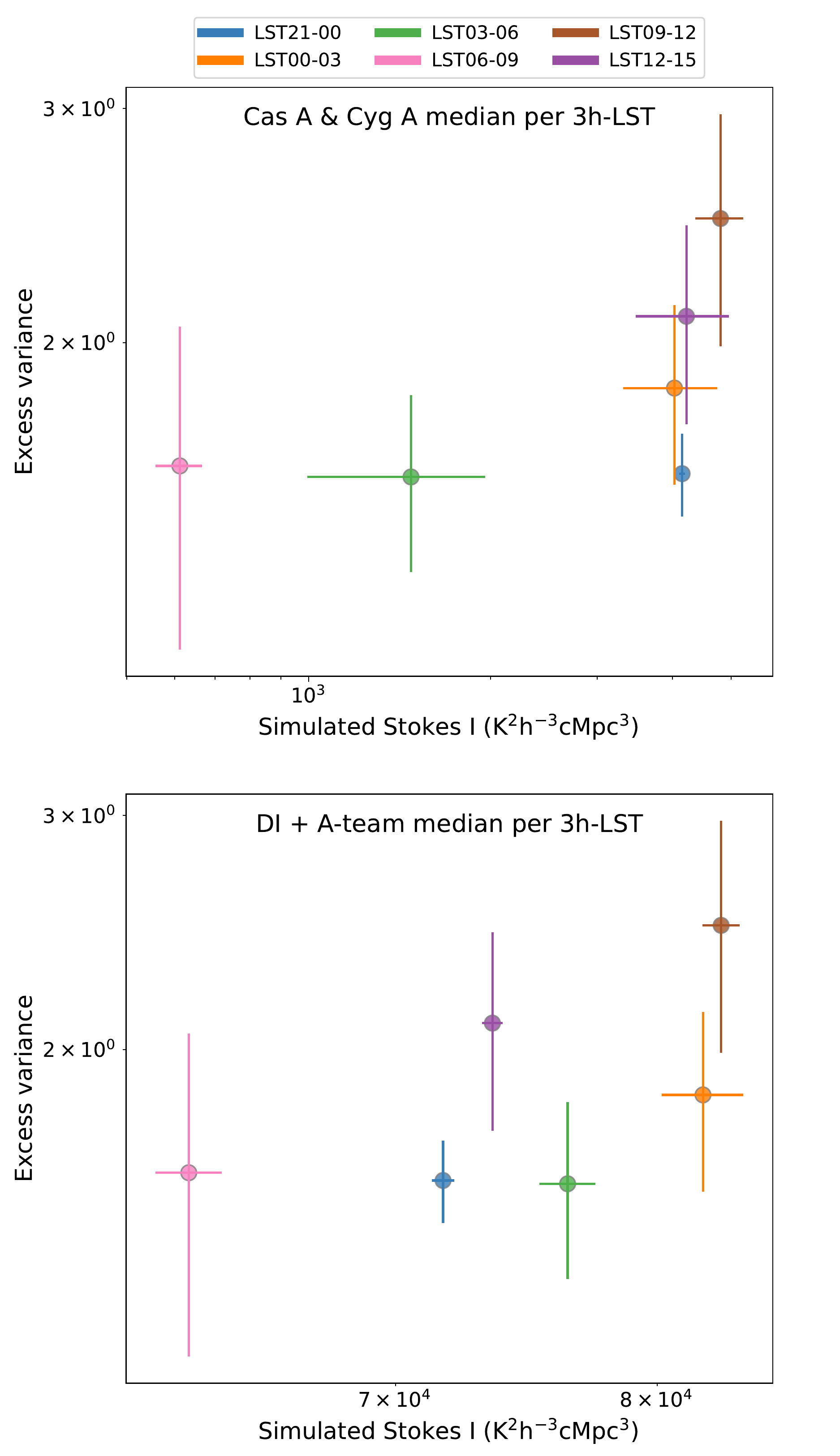}
    \caption{Correlation between the median simulated Stokes I from two sky models and the median excess variance over LST. {\bf Top:} Correlation between the Cas A and Cyg A model and excess variance over LST. {\bf Bottom:} Correlation between the DI sky model (that accounts for $\sim99\%$ of the power in the sky model) and A-team sources and excess variance over LST. Different colors indicate different LST bins. The error bars on each data point indicate the RMS of the simulated Stokes I power and the excess variance between nights per LST bin, respectively (we note that they are not observation errors), to show the night-to-night spread of the data per LST bin. The correlation between the simulated power from the two model and the excess variance show a moderate correlation but it is not strong, due to the limited sample size and night-to-night spread of the data. }
    \label{fig:excess_vs_simu}
\end{figure}
To verify the strength of correlations between the simulated Stokes I power and the excess variance level more accurately, we performed three different correlation tests on the data sets, including a Pearson correlation test as well as Spearman and Kendall rank correlation tests. As we discussed already, the night-to-night difference within a given 3h LST bin can still be significant both for observations and simulations, due to the differences in the number of subbands and flagging status between nights. When the simulated Stokes I and the excess variance level are correlated, this night-to-night difference can possibly introduce outliers and the correlation test results will not reflect the actual strength of the correlation accurately. Thus, in Fig.~\ref{fig:excess_vs_simu}, we show how we took the median value per 3h LST bin for the excess variance and the simulated Stokes I between nights and performed correlation tests on the median values. The rms of the simulated Stokes I power and the excess variance between nights per LST bin are also present on top of each data point, respectively, to show the night-to-night spread per LST bin. 
\begin{table}
\caption{Three correlation test results on the simulated Stokes I power from two sky models and the excess variance level.}
\centering
\begin{tabular}{l|c|c} 
\multicolumn{3}{c}{\textbf{Cas A and Cyg A}}\\ [.5ex] 
\hline
\noalign{\smallskip}
Test type & Correlation coefficient & p-value \\ \hline
\noalign{\smallskip}
Pearson (r) & 0.60 & 0.21 \\
Spearman ($\rho$) & 0.71 & 0.11 \\
Kendall ($\tau$) & 0.60 & 0.14 \\
\noalign{\smallskip}
\hline 
\noalign{\smallskip}
\end{tabular}
\end{table}

\begin{table}
\centering
\begin{tabular}{l|c|c} 
\multicolumn{3}{c}{\textbf{The DI sky model and A-team}}\\ [.5ex] 
\hline
\noalign{\smallskip}
Test type & Correlation coefficient & p-value \\ \hline
\noalign{\smallskip}
Pearson (r) &   0.58 &  0.22 \\
Spearman ($\rho$) &     0.49 &  0.33 \\
Kendall ($\tau$)&       0.33 &  0.47 \\
\noalign{\smallskip}
\hline 
\noalign{\smallskip}
\end{tabular}
\label{tab:correlations}
\end{table}
We performed three types of tests because they are sensitive to different correlation characteristics. The Pearson test is sensitive to linear correlations and is susceptible to outliers. The Spearman and Kendall tests are less sensitive to outliers and are able to detect higher-order monotonic relationships. The correlation test results are summarized in Table.~\ref{tab:correlations}. Overall, the Cas A and Cyg A simulation shows a strong correlation with the excess variance with the correlation coefficients ranging between 0.60 and 0.71, while the DI sky model and A-team simulation shows a rather moderate correlation with the correlation coefficients ranging between 0.33 and 0.58. However, the p-values from both models are relatively high (0.11-0.21 for the Cas A and Cyg A model and 0.22-0.47 for the DI sky model and A-team) and, hence, the correlations are not robust. This is partially due to the limited sample size after processing and the night-to-night spread in each LST bin. These limitations can be mitigated when more nights are combined. At this moment, we conclude that the excess variance correlates moderately with the bright sources in the sky, especially with very strong sources such as Cas A and Cyg A. This indicates that the excess variance might be leakage from the sky power.

\subsubsection{Variance maps}
\label{subsub:sky_image}
\begin{figure*}
    \centering
    \includegraphics[width=\textwidth]{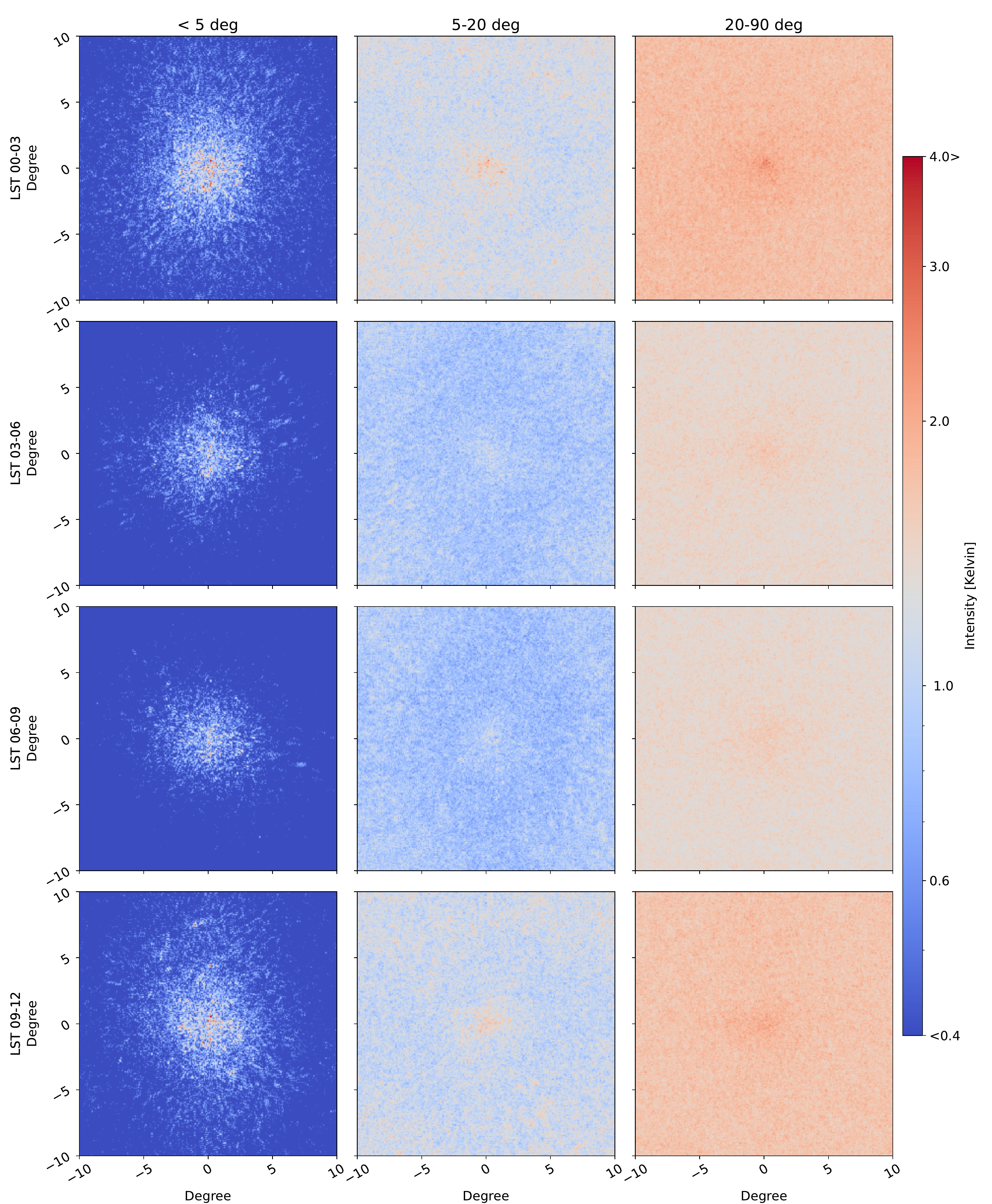}
    \caption{Standard deviation of the residual Stokes I intensities along the frequency direction, after sky-model subtraction and GPR residual foreground removal at 2.5 arcmin resolution incoherently averaged over seven nights per 3h LST from different regions of ($k_\perp$, $k_{||}$) space --- passing only the data under 5$^{\circ}$ delay line (first column), between 5$^{\circ}$ and 20$^{\circ}$ delay lines (second column), between 20$^{\circ}$ and 90$^{\circ}$ delay lines (third column) in the power spectra, respectively. All images are in units of Kelvin. As we show  in Fig.~\ref{fig:13night_2dPS} LST03-06 and LST06-09 once again show lower residual power compared to other LST bins.}
    \label{fig:sky_RMS}
\end{figure*}
\begin{figure*}
    \centering
    \includegraphics[width=\textwidth]{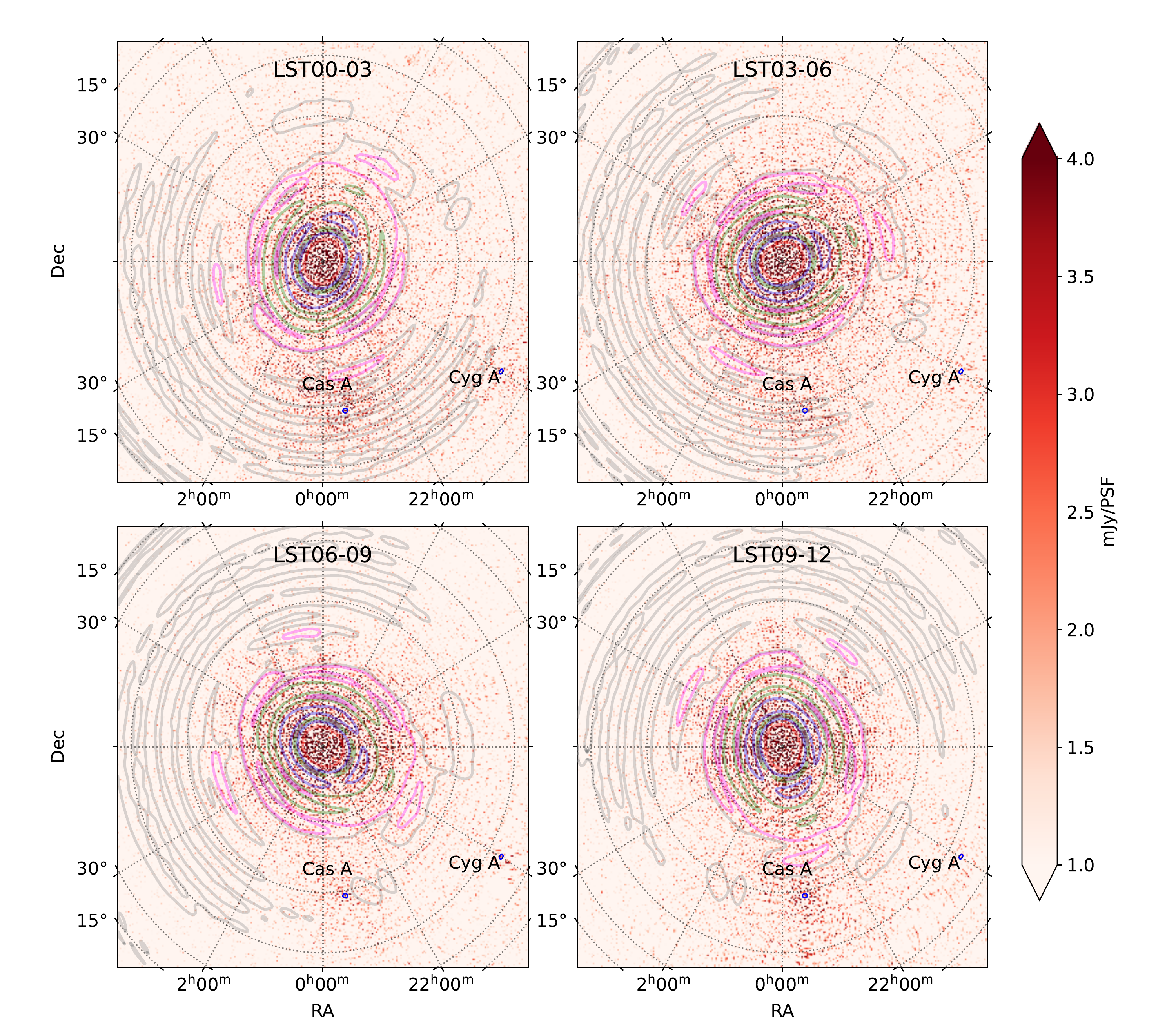}
    \caption{Very wide images after DD-calibration from observation L254116 combining 62 sub-bands with the position of Cas A and Cyg A and the time- and frequency-averaged beam patterns overlaid. From inside to outside, different color (red, blue, green, magenta, and gray) indicate different beam gains, corresponding to $-10$ dB, $-20$ dB, $-30$ dB, $-40$ dB, and $-50$ dB. The actual excess variance we see in the power spectra should be the multiplication of the residual Stokes I power and the beam effects. In the sky images, we see the extra power coming especially from the Cas A direction. This power is at its peak in LST00-03, decreasing in  LST06-09 and increasing again in LST09-12. This tendency corresponds with what we observed in the LST progression of excess variance and the simulated power of Cas A and Cyg A in the top and middle panels of Fig.~\ref{fig:excess_vs_LST}. The averaged beam rotates over time, as we can see in the images, the $-50$ dB sidelobe (in gray) sweeps through Cas A and Cyg A from LST00-03 to LST09-12. In the first three LST slices, the sidelobes rotate away from Cas A and Cyg A, at the same time, the power from Cas A and Cyg A decreases. As a result, the excess variance effect from Cas A and Cyg A decreases from LST00-03 to LST06-09. In LST09-12, although the $-50$ dB envelope is located in the opposite direction of Cas A and Cyg A, the $-40$ dB sidelobe (in magenta) has an extended arm towards Cas A direction and the extra power from Cas A direction increases again. This combined effect results in increased power in LST09-12, as we see in Fig.~\ref{fig:excess_vs_LST}.}
    \label{fig:very_wide}
\end{figure*}
We created wide-field (20$^{\circ}\times$20$^{\circ}$) maps of the residual Stokes I intensity standard deviation along frequency incoherently averaged over seven observations to assess whether the excess variance in the power spectra comes from specific sources or directions in the sky. These images are created after DI-, DD-calibration and GPR foreground removal from seven observations, focusing on the excess variance dominant part of the power spectra (see Fig.~\ref{fig:13night_2dPS}), under the instrument horizon. For each 3h LST residual Stokes I power spectrum, we create maps from specific regions in the power spectra --- passing only the data under 5$^{\circ}$ delay line (first column; roughly the primary beam area), between 5$^{\circ}$ and 20$^{\circ}$ delay lines (second column), between 20$^{\circ}$ and 90$^{\circ}$ delay lines (third column), respectively. The filtered power spectra are Fourier transformed along $k_\text{x}$, $k_\text{y}$ ($k_\perp = k_\text{x}^2 + k_\text{y}^2$), and $k_{||}$ directions into image space. The created images are shown in Fig.~\ref{fig:sky_RMS}. 

In all wide-field maps, the residual power is stronger around the central part where the instrument has the highest sensitivity and the power spectra are estimated (on the central $4^{\circ}\times4^{\circ}$ FoV). For a fixed LST bin, the residual power from the primary beam region is mostly added to the central part of images (first column) and the residual power outside the primary beam is scattered more evenly over the entire FoV (second and third columns). From left to right, we plot the images corresponding to the different regions of the power spectra in ($k_\perp$, $k_{||}$) space, however, the power difference is marginal and we see a plateau of the residual power in the EoR window (third column). This indicates that most of the residual power still comes from foregrounds (first two columns), and some of this power is not from local sources (that they would add power as single sources as shown in the first column), but possibly from distant sources such as Cas A and Cyg A (which scatter power rather evenly over the entire FoV). Even after the calibration methods including DI-, DD-calibration, sky-model subtraction and GPR foreground removal, these foreground sources are not sufficiently removed. Therefore, further improvement of calibration methods, especially on foreground removal is still needed.

In Fig.~\ref{fig:very_wide}, we present very wide-field images ($100^\circ\times100^\circ$) after DD-calibration to investigate how the power from Cas A and Cyg A contributes to the excess variance. The images are created from observation L254116 combining 62 sub-bands and applying the baseline cut $50-250\lambda$ used for the power spectrum estimation. The positions of Cas A and Cyg A are marked with blue circles on top. The time- and frequency-averaged beam patterns per 3h LST are also overlaid on top with different colors indicating different gains (from inside to outside: red, blue, green, magenta, and gray corresponding to $-10$ dB, $-20$ dB, $-30$ dB, $-40$ dB, and $-50$ dB gains, respectively).

The actual residual power we see in the power spectrum is the multiplication of the sky power in the images and the average beam. The residual power in the very wide-field sky images in Fig.~\ref{fig:very_wide} show that the residual power sharply decreases outside the primary beam (the LOFAR core station FWHM at 150 MHz is around $3.8^\circ$~\citep{van_Haarlem_2013_refId0}), but there is extra power coming from the direction of Cas A and Cyg A, the power is especially strong from the Cas A direction. The power changes over LST and becomes minimum at LST06-09. This is in line with the excess variance and the simulated Cas A and Cyg A power progression over LST in Fig.~\ref{fig:excess_vs_LST}. 
The averaged beam pattern rotates over time and the $-50$ dB sidelobe (in gray) sweeps through Cas A and Cyg A, the overlap is maximum in LST00-03 and minimum in LST09-12. This explains the decreasing excess variance in the first three LST slices. In LST09-12, the residual power in the sky from Cas A is strong again. While the $-50$ dB sidelobe is almost on the opposite side of Cas A and Cyg A this time, however, the $-40$ dB sidelobe (in magenta) has an extended arm towards Cas A direction that allows power from Cas A to enter. The combination of these two effects possibly results in a higher excess variance in Fig.~\ref{fig:excess_vs_LST}. 

\begin{figure}
    \centering
    \includegraphics[width=250px]{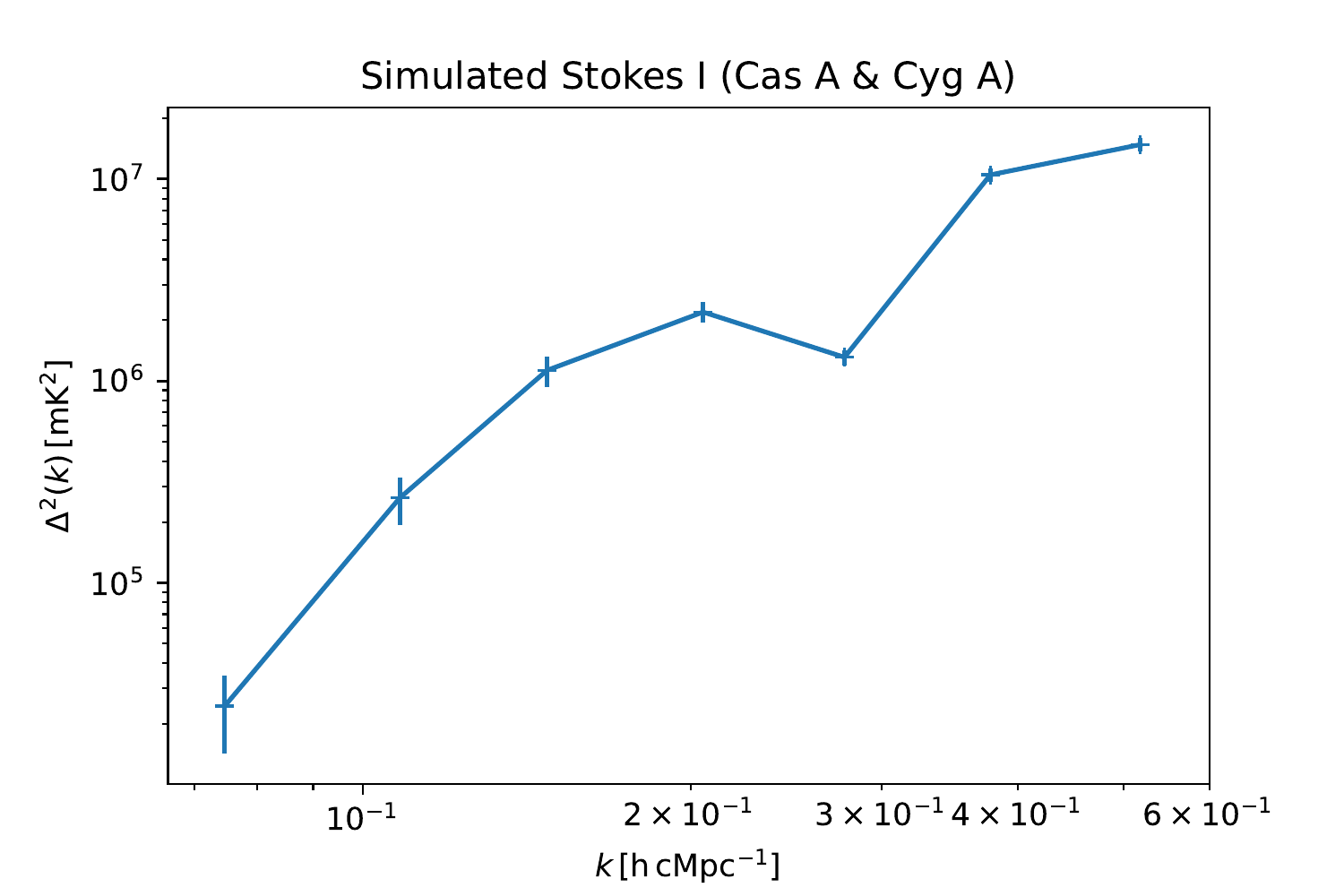}
    \caption{Spherically averaged power spectrum estimated from one-night (L246309) Cas A and Cyg A only simulation. Compared the Cas A and Cyg A introduced Stokes I power to the estimated 21 cm signal at $k=0.1\text{ }hc\text{Mpc}^{-1}$, the Cas A and Cyg A power should be reduced by $\sim99.3\%$ or even higher in the calibration to be able to distinguish the 21 cm signal from the Cas A and Cyg A power. }
    \label{fig:3dps_cc}
\end{figure}

This further supports that the excess power depends on LST and that it is correlated with sky-related effects, such as gain errors from an imperfect sky model or extra power from the sky that is not part of the sky model. These effects are not corrected in the DD calibration. Using the simulated Stokes I power of Cas A and Cyg A as an example, we estimate the calibration accuracy required to avoid the excess variance from the two sources. The spherically averaged power spectrum of Cas A and Cyg A from one-night observation (L246309) is shown in Fig.~\ref{fig:3dps_cc}. The simulated power from only Cas A and Cyg A observed by the LOFAR-HBA system is around $2\cdot10^5\text{ mK}^2$ at $k=0.1\text{ }hc\text{Mpc}^{-1}$, while the estimated power from the 21 cm EoR signal is around $10\text{ mK}^2$ level at $k=0.1\text{ }hc\text{Mpc}^{-1}$. Based on these results, we estimate that the Stokes I power of Cas A and Cyg A must be reduced at least by $\sim99.3\%$ to be able to distinguish the 21 cm signal from the excess variance caused by Cas A and Cyg A. 

\subsection{Ionospheric effects}
\begin{figure*}
\centering
\textbf{Diffractive scale $r_\text{diff}$ and excess variance correlation\\}
    \includegraphics[width=420px]{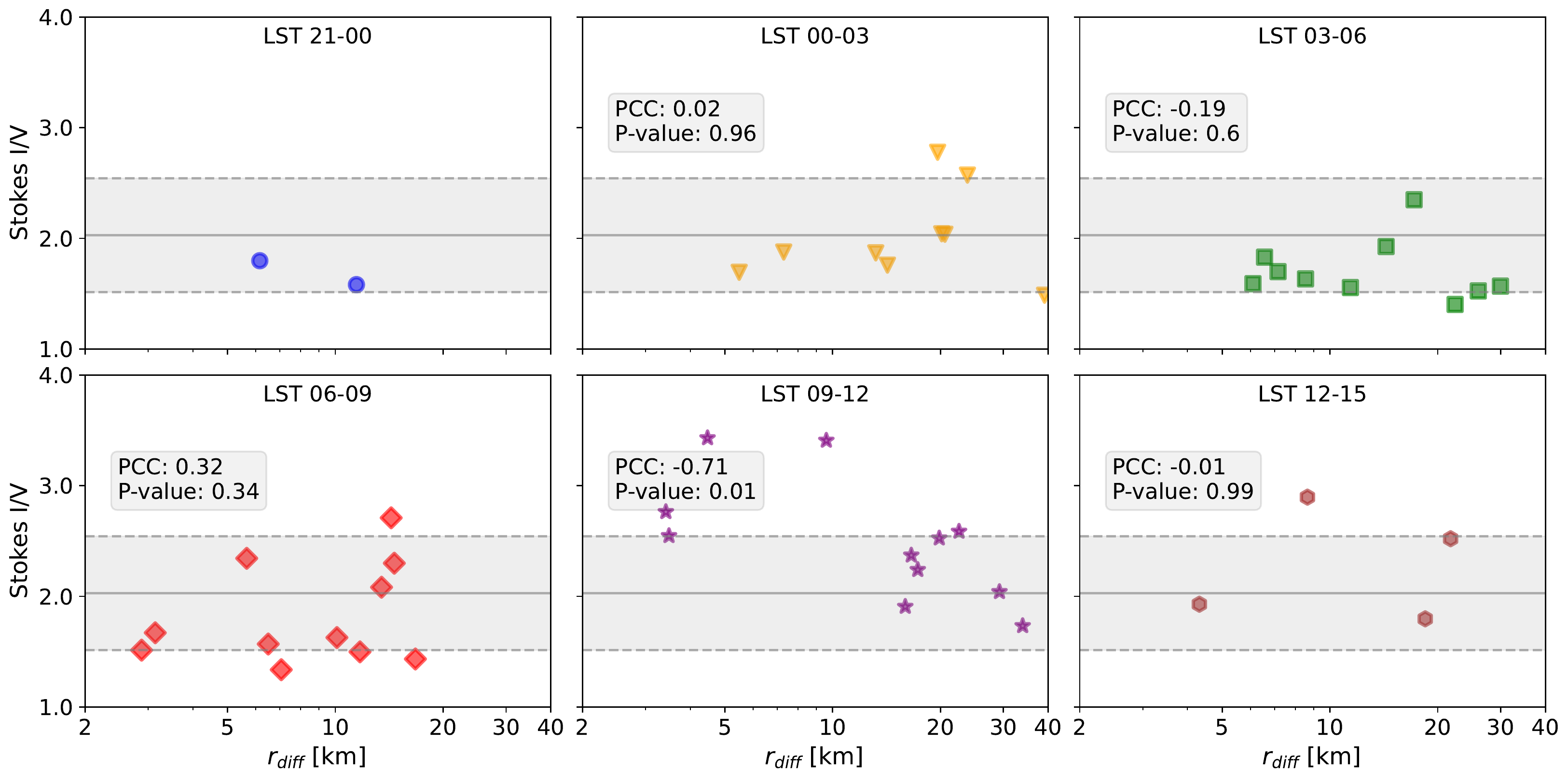}
\textbf{\\Structure function fitted slope $\beta$ and excess variance correlation\\}
    \includegraphics[width=420px]{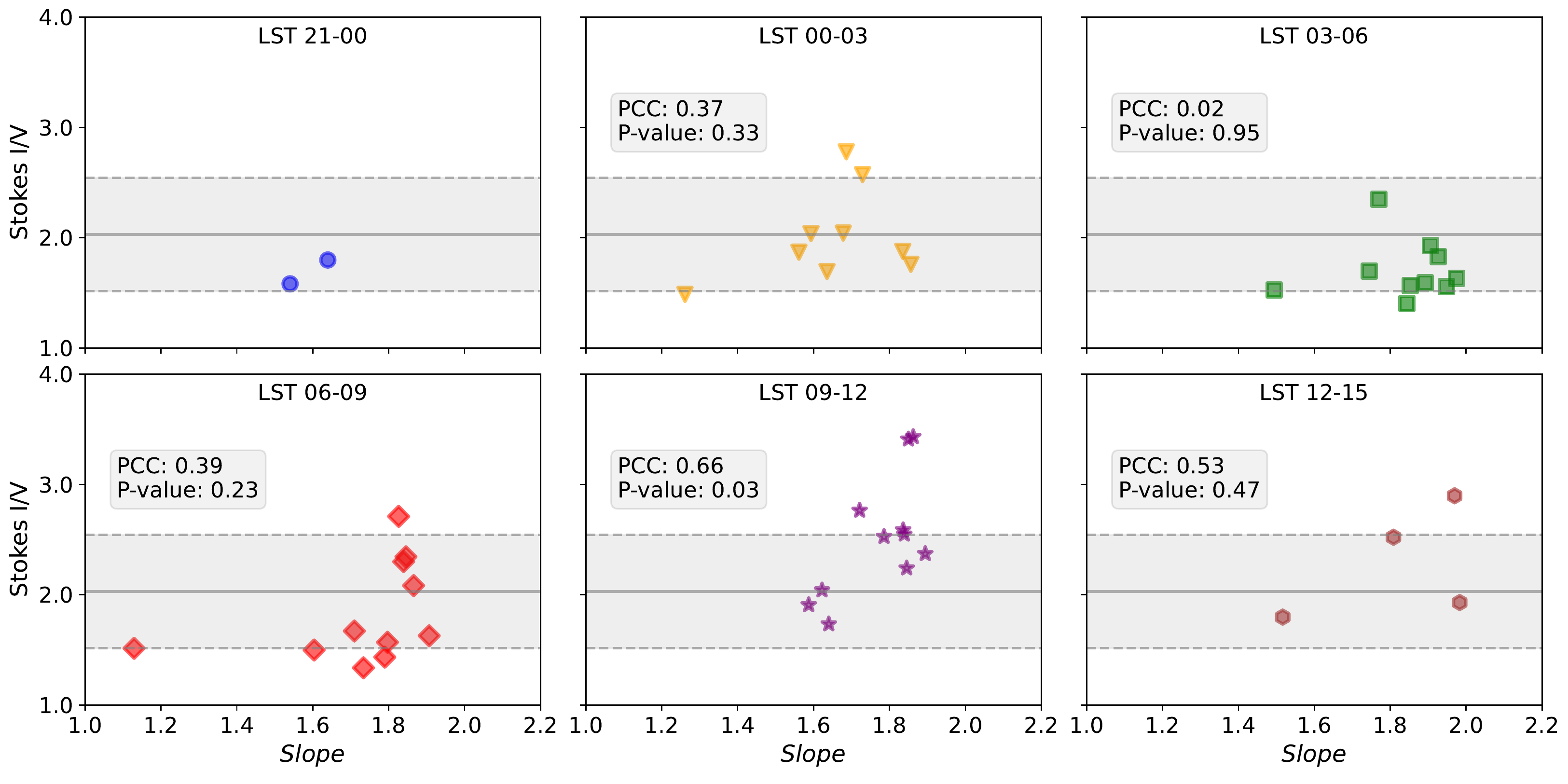}
    \caption{Excess variance (i.e., the Stokes I and V ratio) as a function of diffractive scale $r_\text{diff}$ (top two rows) and fitted slope $\beta$ (bottom two rows) at 140 MHz per LST bin for 13 observations with the shaded area indicating the $1-\sigma$ excess variance over 13 observations and all LST bins. The color denotes different LST bins. The ionosphere and excess variance distributions vary from night to night (different data points per LST bin) and also from LST to LST. No strong correlation between the excess variance and the diffractive scale is found at the current level.}
    \label{fig:iono_excess}
\end{figure*}
In this subsection, we investigate potential correlations between the excess variance and ionospheric effects using the phase structure function, from which we derive the fitted slope $\beta$ and the diffractive scale, $r_\text{diff}$. These metrics are used to describe ionospheric conditions during observation and are obtained from LOFAR observations by fitting the phase variance as a function of baseline length per 3h LST bin (we refer to~\cite{2016RaSc...51..927M_Mevius_2016}, Section.~\ref{sec_sources} and Appendix~\ref{appendix} for details). 
Figure~\ref{fig:structure_fn} shows the binned phase structure functions of 13 observations at 140 MHz for LST00-03 (top three rows) and LST03-06 (bottom three rows), with the results of the power-law phase-structure function fits superimposed with solid lines in red. We calculate the Pearson's coefficient $\text{R}^2$ to estimate the quality of the fit. $\text{R}^2$ ranges between 0.87 and 0.99 for all observations and LST bins, yielding representative $\beta$ and $r_\text{diff}$ values. We list the resulting values $\beta$, $r_\text{diff}$ and $\text{R}^2$ per LST bin for 13 observations in Table.~\ref{table:ionosphere}. 

Based on the study from \cite{2016RaSc...51..927M_Mevius_2016}, we expected the fitted slope $\beta$ from structure functions to range between 5/3 (a contribution from the pure Kolmogorov turbulence in the lower atmosphere, \cite{053a1363370540658ce7bd7cc38617ad_van_Velthoven}) and 2.0 (from non-turbulent structures, such as travelling ionospheric disturbances (TIDs), \cite{https://doi.org/10.1029/JA077i025p04761_Rufenach}). From our results, about 73\% of $\beta$ values correspond to this range. Furthermore, \cite{10.1093/mnras/stv1594} showed that a minimum diffractive scale of 5 km at 150 MHz is required for an EoR detection. About 87\% of $r_\text{diff}$ is above 5 km for all LST bins and 13 observations, which is encouraging.  

Finally, we correlated $\beta$ and $r_\text{diff}$ with the excess variance per LST bin, respectively, to assess whether the ionosphere has a sky-dependence or correlates with the excess variance. The results are shown in Fig.~\ref{fig:iono_excess}. In addition, we performed a Pearson correlation test per LST bin to provide a PCC and a p-value as single number measures for the correlation between the ionosphere and excess variance. The results are summarized in Table.~\ref{table:ionosphere}. The PCC values suggest no strong correlations (PCC $<0.6$) between the ionosphere and excess variance over different LST bins, neither for $r_\text{diff}$ nor $\beta$, with a possible exception for the bin LST09-12. We expect that ionospheric effects are more likely to correlate with Earth-based effects than with the sky, thus, they are more correlated with UT than with LST. If there is any correlation between the ionosphere and excess variance, this should appear consistently over all LST bins. The PCCs between LST bins vary widely in our results, ranging between -0.71 and 0.32 for $r_\text{diff}$ and between 0.02 and 0.66 for $\beta$, respectively. In particular for $r_\text{diff}$, the PCC alternates between positive and negative correlations for different LST bins. We suspect that these slightly higher PCCs in LST09-12 (-0.71 for $r_\text{diff}$ and 0.66 for $\beta$) more likely come  from a random statistical error than a real correlation. In addition, our samples per LST bin are very limited (from 2 to 11 per bin, see Fig.~\ref{fig:LST_slice}), and having one or two noisy data points in a bin could heavily affect the final result of the Pearson test. In summary, we conclude that the excess variance is not strongly correlated with the ionosphere at the current level. 


\section{Conclusions}
\label{sec:conclusions}
The LOFAR-EoR Key Science Project is aimed at detecting the 21 cm hydrogen signal originating from the Epoch of Reionization in the redshift range of $z\approx7-11$. The detection of the 21 cm signal, however, requires very high precision calibration to correct corrupting effects from the strong galactic and extragalactic foregrounds, the ionosphere, RFI and instrument's limitations, and so on. After the current best-effort in calibration, there is still extra power that is above the expected thermal noise level in the 21 cm power spectra,  which is known as the "excess variance"~\citep{10.1093/mnras/stw2277_patil2016,Patil_2017,2020MNRAS.493.1662M_Mertens_2020}. There are two known types of this excess variance: (1) the spectrally-uncorrelated excess on short baselines ($< 100\lambda$) and (2) the spectrally and temporally (within a few nights) correlated excess ($l_\text{ex}\sim0.25-0.45$ MHz) that is stronger in the foreground dominated wedge~\citep{2020MNRAS.493.1662M_Mertens_2020}. In this paper, we examine the correlations between possible sources, such as errors resulting from DD-gain solutions, sky-related effects, and ionospheric effects as well as the excess variance. Our two main conclusions are as follows: 

We find no correlation between either the diffractive scale or the ionospheric phase structure function fit slope and excess variance,  suggesting that the ionosphere is not a dominant contributor to the excess variance at the current level of sensitivity.    

On the other hand, the level of excess variance shows a trend with local sidereal time (LST), suggesting that excess variance depends on the sky as affected by the instrument (not the ionosphere). By limiting observations into equivalent 3h LST bins, the difference in the point spread functions (PSFs) and primary beams between nights is minimised within each LST bin. In this way, sky-related effects are closely related to PSFs and primary beams and, hence, they will be reflected as an LST dependence. Investigating this further, we find that the simulated Stokes I power of bright sources in the sky model shows a similar progression in LST with the excess variance. 

We show that the excess variance presents a similar LST dependence in the residual Stokes I images. The images show a clear imprint of the primary beam (within $5^\circ$ from the phase centre), but also a plateau for the power spectrum region in the EoR window. In our very-wide sky images, we further show that some of the power left after DD-calibration is coming from the direction of Cas A and Cyg A. These results again support the conclusion that excess variance could be leakage from the sky power, and some of this power is not the result of local sources (in the case of which, there should be no plateau), but of distant sources such as Cas A and Cyg A.

We conclude that excess variance, as seen on the NCP with the LOFAR-HBA system, is likely dominated by residuals of bright sources after sky-model subtraction due to their still-limited sky models, which is further exacerbated by flagging and gridding effects. Although it has been suspected that the ionospheric effects might significantly contribute to the residual noise in the 21 cm power spectra, we conclude that the ionosphere does not show any strong correlation with the excess variance at the current level. Rather, the excess variance is correlated with LST and the sky model. Thus our future analysis will focus on analyzing the time dependence of gain solutions, enforcing smoother solutions in time, and improving the sky model. 

Based on our analysis, we suggest the following strategies for future improvements:

{\bf Improving the sky model:} the correlation between the excess variance and the sky model is strong. We know that the current sky model is not perfect and this incompleteness in the model can be further propagated to gains during calibration and further contribute to the excess variance in the power spectra. Our current sky model does not include diffuse Stokes I, Q, and U emission and the inclusion of the diffuse components will improve the power spectrum results \citep{prasad2016aartfaac,99e589c5e759487a99144c9f65b4ff95_Gehlot}.

{\bf Applying finer time intervals for DD-gains:} as we see in Subsection.~\ref{subsubsec:gain_smoothness}, the smoothness of DD-gains in time is not as good as the one in frequency. Depending on the cluster, we still see jumps between time-consecutive gains, as in Cluster82 in Fig.~\ref{fig:f_avg}. We suspect that for some clusters, the current solution intervals are too long to take into account some rapidly-varying ionospheric scintillation whose decorrelation time is a few seconds~\citep{10.1093/mnras/stw443_Vedantham2016}. Solving gains for shorter time intervals can help enforce smoother gains in this case.   

{\bf Focusing on a "cleaner" time window:} we see both from the observations and simulations that the excess variance and the power from foreground sources are lower in a specific time range, LST03-09. While the excess variance is possibly correlated with errors from the calibration or the incompleteness in the sky model, it is difficult to change the current sky model or the calibration process to fundamentally solve the problem. However, since we now know that the time window LST03-09 is relatively "clean" in terms of the excess variance, we can "avoid" the excess variance by focusing on this specific time range for future analyses.

{\bf Categorizing the excess variance in $k$-space:} based on observations and simulations, we know that most of the power in cylindrical power spectra is concentrated in the foreground wedge (the region below the 20$^{\circ}$ delay line in the power spectra). The excess variance is also higher in the wedge region. Our current approach calculates the excess variance per 3h LST bin by averaging over all $k$ values and this does not take into account the $k$-dependence nature of the excess variance. In a future work, we plan to categorize the excess variance depending on $k$. This will better address the correlation between the excess variance and its possible causes. 

{\bf Combining more data:} as we discussed in Sec.~\ref{subsec:sky_effects}, the correlations between the simulated Stokes I from sky models and excess variance are non-negligible but not yet robust, given high p-values from correlation test results. This is partially due to the limited sample size and the night-to-night variation in each LST bin. By combining a greater number of nights, the night-to-night variation in each LST bin will be reduced and the correlation will become more clear (if a correlation indeed exists).

The analysis in this paper shows that the excess variance is more likely to be correlated with the sky-related effects than other effects such as ionospheric effects. In particular, the far and bright sources such as Cas A and Cyg A dominate the effects. The correlation, however, is not yet decisive. By implementing the proposed strategies, we will be able to reduce the level of excess variance and reach deeper limits in the near future. 

\begin{acknowledgements}
       The authors would like to thank the anonymous referee and the editor for the comments which helped improve and clarify this manuscript. HG and LVEK would like to acknowledge support from the Centre for Data Science and Systems Complexity (DSSC) at the University of Groningen and a Marie Sk\l{}odowska-Curie COFUND grant, no.754315. AG would like to acknowledge IUCAA, Pune for providing support through the associateship programme. BKG gratefully acknowledges the support from the National Science Foundation through award for HERA (AST-1836019). ITI was supported by the Science and Technology Facilities Council [grant numbers ST/I000976/1 and ST/T000473/1] and the Southeast Physics Network (SEP-Net).
\end{acknowledgements}

\section*{Data Availability}

The data underlying this article will be shared on reasonable request to the corresponding author.
%
%
\bibliographystyle{aa} 
\bibliography{ref}
\begin{appendix} 
\section{Impact of flagging on simulated data}
\label{appendix2}
The visibility simulations performed in this work are based on real observations. This implies that the simulations will have the equivalent RFI flagging (on baselines) and $uv$-coverage as observations. We notice that these can be a disadvantage for the power spectrum estimation of the simulated data: (1) simulations are not contaminated by RFI, hence having extra baselines flagged will lead to a data loss and this might introduce extra power in the power spectra~\citep{10.1093/mnras/stz175_offringa_2019a}; (2) when simulations are sliced into 3h LST bins, the $uv$-coverage will be limited, as we have seen with the observations. Hence, it may be necessary to flag $uv$-cells with bad coverage even in the simulations before the power spectrum estimation to optimize the results. 

\begin{figure*}
    \centering
    \includegraphics[width=250px]{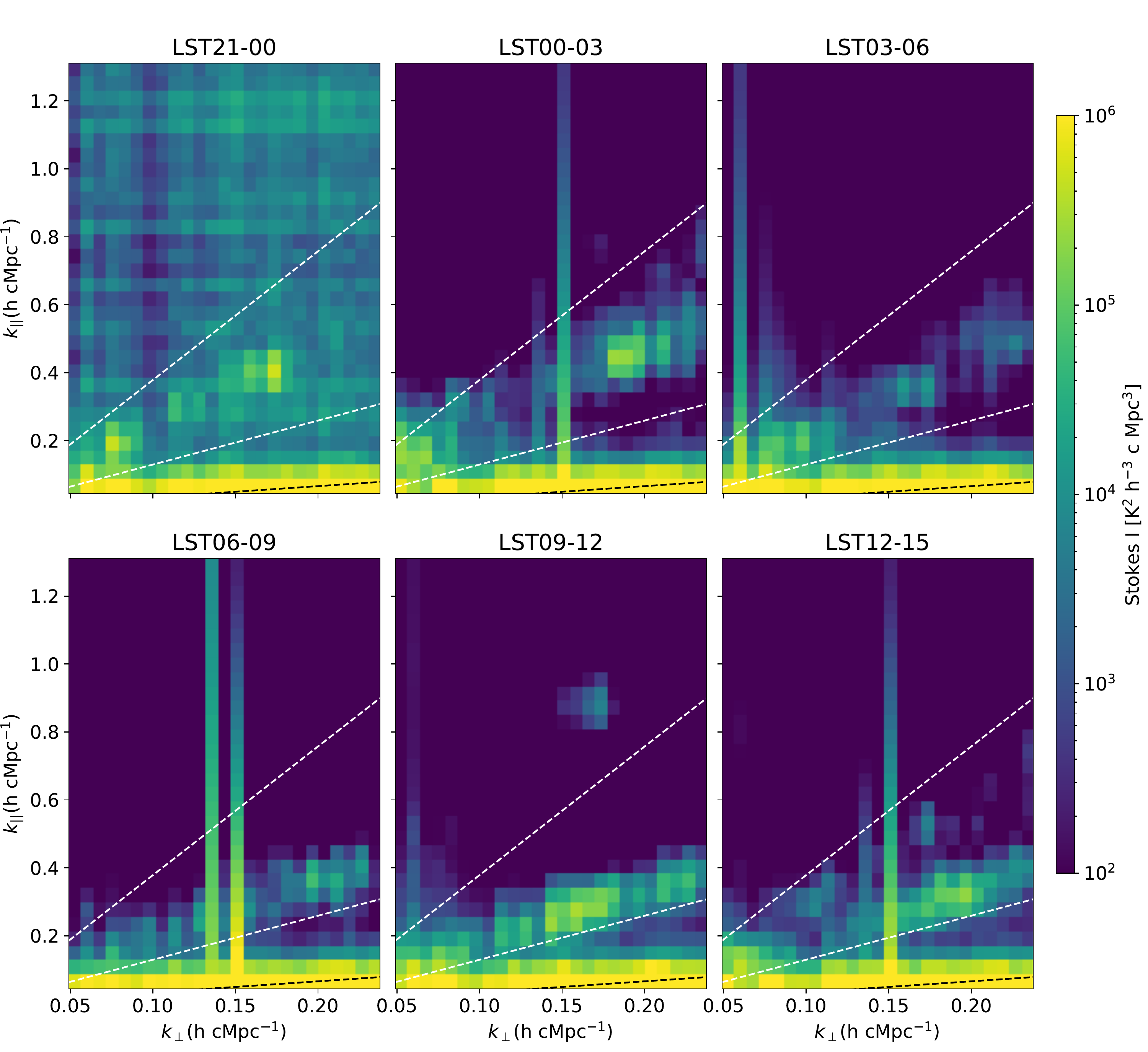}
    \includegraphics[width=250px]{plots/simu_di_flag2_median_fov4_v3.pdf}\\
    \includegraphics[width=250px]{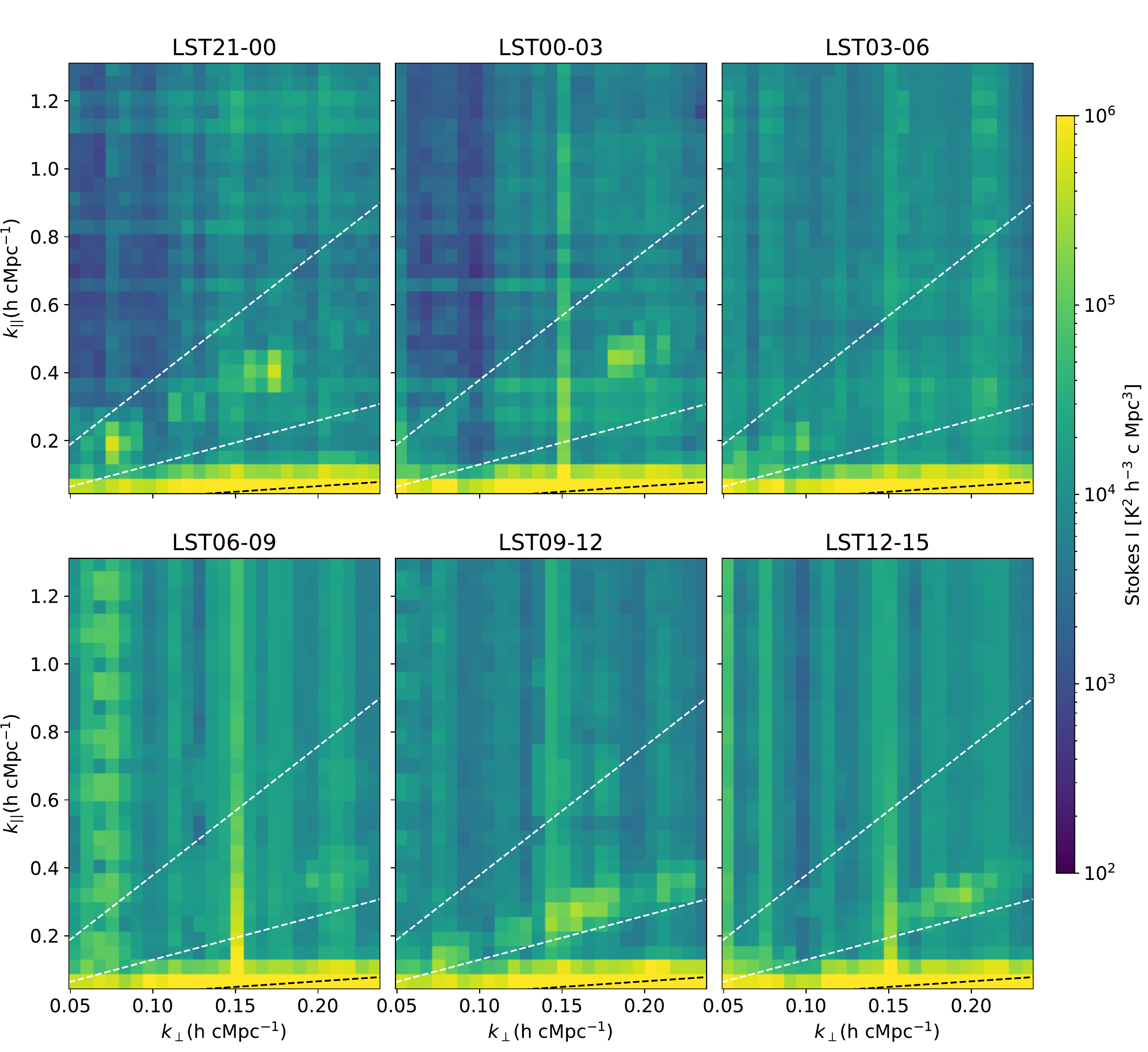}
    \includegraphics[width=250px]{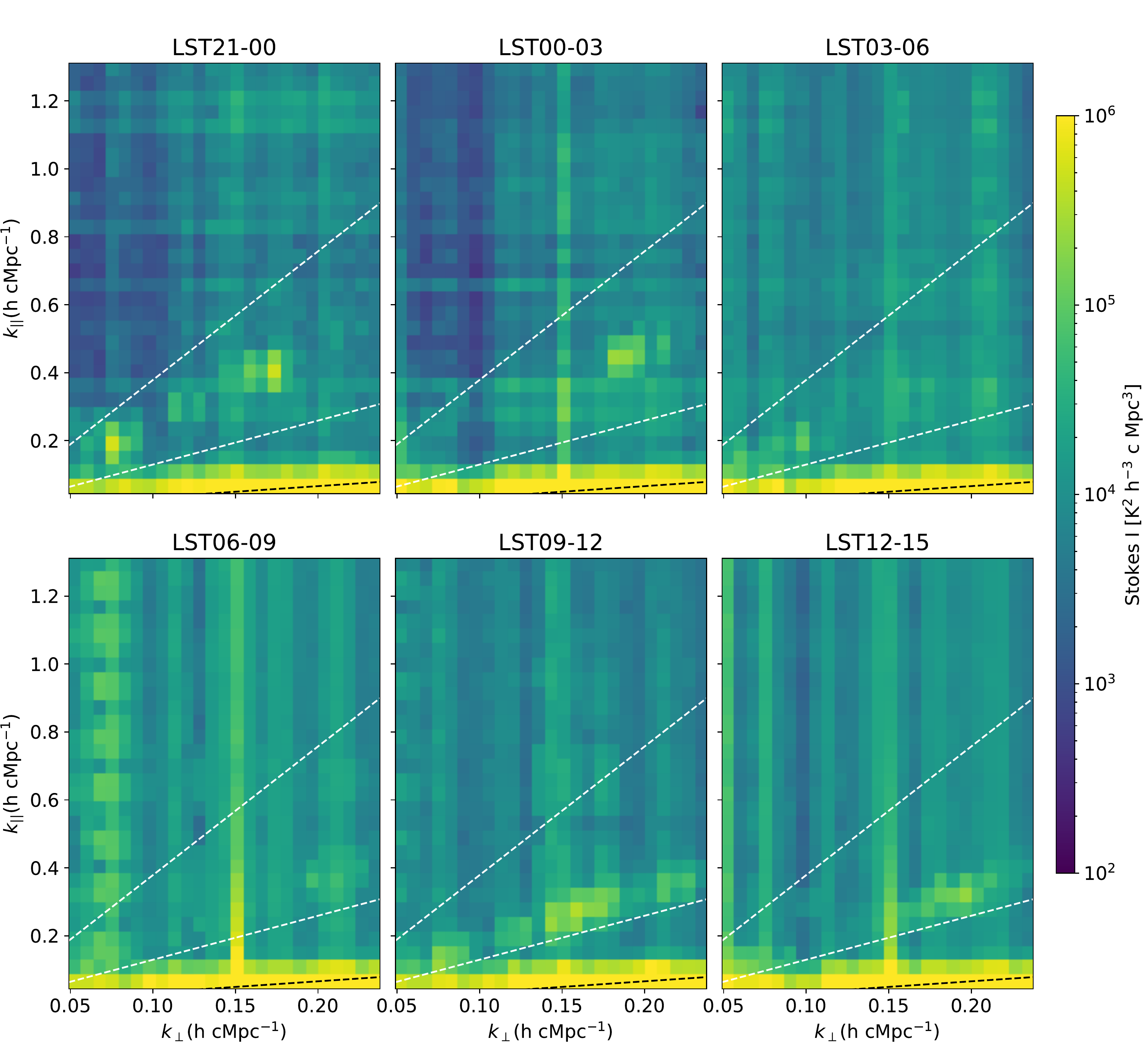}
    \caption{Median cylindrical Stokes I power spectra of the 13 night simulations with the DI-calibration model (including the 1416 brightest components that account for $\sim99\%$ of the flux from our sky model) and A-team sources per LST bin with 4 different flagging conditions. {\bf Top left:} $uv$-cells not flagged and RFI unflagged. {\bf Top right:} $uv$-cells flagged and RFI unflagged (the condition we use for the analysis). {\bf Bottom left :} $uv$-cells not flagged and RFI flagged. {\bf Bottom right:} $uv$-cells flagged and RFI flagged. Dashed lines indicate, from bottom to top, 5\degr~(the primary beam), 20\degr~and instrumental horizon delay lines. RFI flagging adds extra power to the EoR window (bottom left and right) and unflagged bad $uv$-cells add vertical structures in the power spectra (top left). Hence, applying $uv$-cell flagging and unflagging RFI (top right) is optimal for the power spectrum estimation in this case. }
    \label{fig:2dps_flag_test_di}
\end{figure*}

To investigate the impact of RFI flagging from observations and the poor $uv$-coverage on the simulations, we estimate 3h LST power spectra of simulated visibilities (using the 1416 brightest and A-team sources) applying four different flagging conditions (summarized in Table.~\ref{tab:flags}): (1) $uv$-cells not flagged and RFI unflagged; (2) $uv$-cells flagged and RFI unflagged (the condition we use for the analysis); (3) $uv$-cells not flagged and RFI flagged; (4) $uv$-cells flagged and RFI flagged.
\begin{table}
\caption{Summary of four different flagging conditions used for testing.}
\begin{center}
\begin{tabular}{c|l|l} 
\hline
\noalign{\smallskip}
Condition & $uv$-cell flagging & RFI flagging \\ \hline 
\noalign{\smallskip}
 1 & not applied & unflagged \\
 2 & applied & unflagged \\
 3 & not applied & flagged \\
 4 & applied & flagged \\
 \noalign{\smallskip}
\hline 
\noalign{\smallskip}
\end{tabular}
\label{tab:flags}
\end{center}
\end{table}    

The results of power spectra with different flagging conditions are shown in Fig.~\ref{fig:2dps_flag_test_di}. The power spectra clearly show that having RFI flagging certainly introduces extra power over the EoR window (bottom left and right in Fig.~\ref{fig:2dps_flag_test_di}). Since we are simulating bright sources and their fluxes are known to be smooth in frequency, the resulting power should be constrained in the foreground wedge, especially, below the instrument horizon. Any extra structures, especially above the foreground wedge, are considered to be from effects other than foregrounds. Once the RFI flagging is removed (top left in Fig.~\ref{fig:2dps_flag_test_di}), the added power is significantly reduced. However, we see some vertical structures around $k_{\perp}\sim0.13-0.15 \text{ }hc\text{Mpc}^{-1}$ in LST00-03, LST06-09 and LST12-15, the added power comes from a few $uv$-cells with bad coverage. After the application of $uv$-cell flagging, we note that the vertical power from the bad $uv$-coverage is reduced, while the power from foregrounds remains in the foreground wedge. Therefore, we conclude that applying a $uv$-cell flagging and RFI unflagging (flagging condition 2) is optimal for the power spectrum estimation in this case.

\section{Ionospheric phase structure function metrics} 
\label{appendix}
The spatial structure of the ionosphere can be described by its power spectrum or its Fourier inverse "phase structure function," which is defined as follows~\citep{Tol_2009,2016RaSc...51..927M_Mevius_2016}: 
\begin{equation}
    D(r) = \langle \big( \phi(r') - \phi(r' + r) \big)^2 \rangle,
\end{equation}
where $\phi(r)$ is an ionospheric phase at a position $r$. For Kolmogorov turbulence, the phase structure function is reduced to:
\begin{equation}
    D(r) = \left(  \frac{ r } { r_\text{diff}}  \right) ^ \beta,
\end{equation}
where $r_\text{diff}$ is the spatial scale where the phase variance corresponds to $1 \text{ rad}^2$, which is known as the diffractive scale~\citep{doi:10.1098/rsta.1992.0090_Narayan}. For pure Kolmogorov turbulence, $\beta$ corresponds to 5/3.

\begin{figure*}
\centering
\textbf{\\Structure function fits LST12-15\\}
{\includegraphics[width=400px]{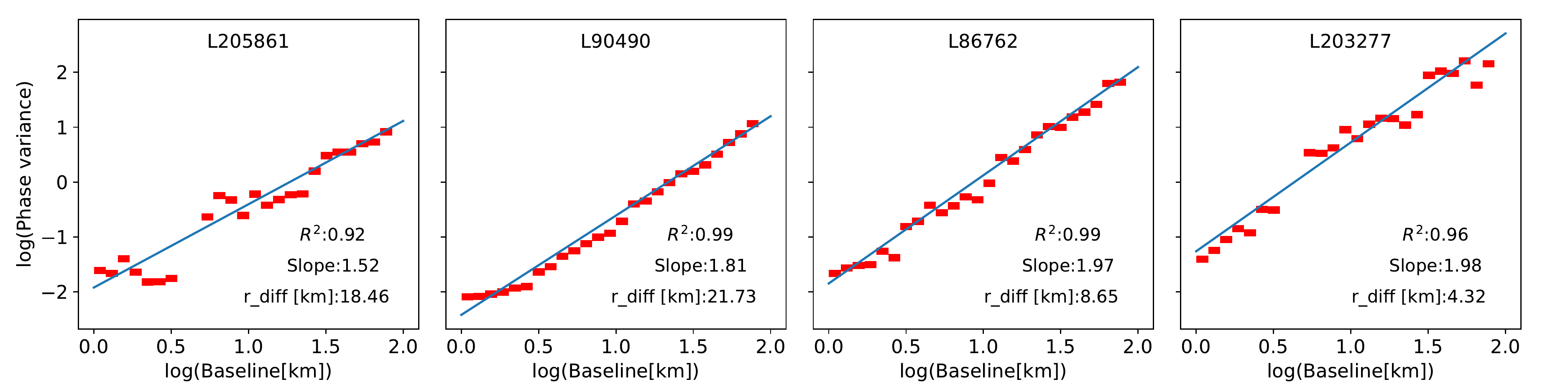}} 
\textbf{\\Structure function fits LST21-00\\}
{\includegraphics[width=220px]{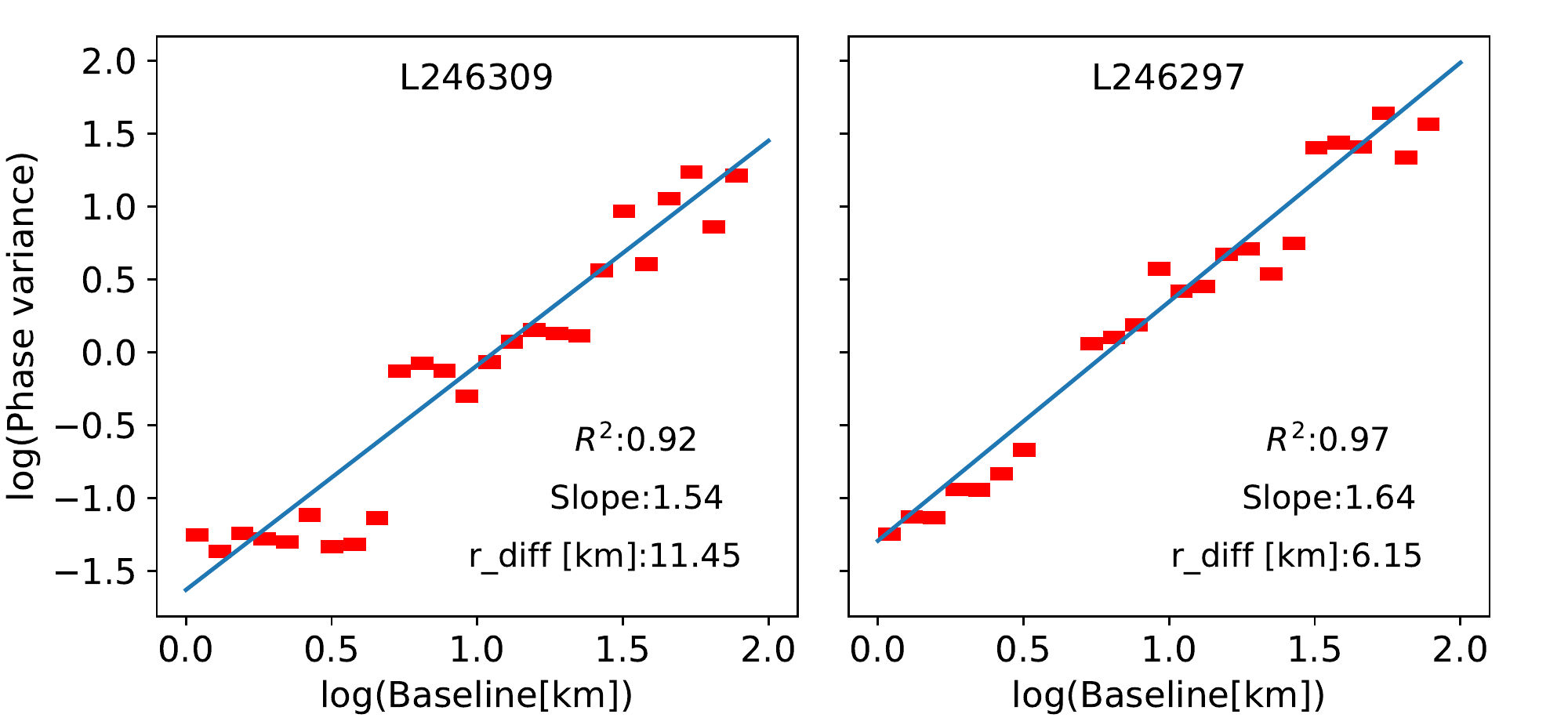}}  \\

\caption{Binned phase structure function fits of thirteen observations at 140 MHz for 3h LST bins. The fitted results are superimposed with solid lines in blue.}
    \label{fig:structure_fn}
\end{figure*}

In particular, $\beta$ and $r_\text{diff}$ are useful single number metrics for the ionospheric quality. We construct the phase structure function for the 13 night observations and fit for parameters $\beta$ and $r_\text{diff}$ to determine the ionospheric quality per night and 3h LST bin (see~\cite{2016RaSc...51..927M_Mevius_2016} for details). The phase structure functions per night and 3h LST are shown in Fig.~\ref{fig:structure_fn} and the results are summarized in Table.~\ref{table:ionosphere} with the coefficient of determination $\text{R}^2$ for each constructed phase structure function to provide the quality of the fit.

\begin{table*}
\centering
\caption{Summary of structure function fitted parameters per LST bin for 13 observations.}
 \begin{tabular}{l r r r r r r} 
 \multicolumn{7}{c}{\textbf{Structure function fitted slope $\beta$}}\\ 
  \noalign{\smallskip}
 \hline \noalign{\smallskip}
 Observation ID & LST21-00 & LST00-03 & LST03-06 & LST06-09 & LST09-12 & LST12-15\\ 
  \noalign{\smallskip}
 \hline \noalign{\smallskip}
 L80847 & - & 1.69 & 1.91 & 1.83 & 1.85 & -\\
 L80850 & - & 1.73 & 1.85 & 1.87 & 1.89 & -\\
 L86762 & - & - & 1.77 & 1.85 & 1.86 & 1.97 \\
 L90490 & - & - & - & 1.84 & 1.85 & 1.81 \\
 L196421 & - & 1.56 & 1.84 & 1.79 & 1.64 & -\\
 L203277 & - & - & - & 1.73 & 1.72 & 1.98 \\
 L205861 & - & - & - & 1.13 & 1.59 & 1.52 \\
 L246297 & 1.64 & 1.64 & 1.89 & - & - & -\\
 L246309 & 1.54 & 1.26 & 1.49 & - & - & -\\
 L253987 & - & 1.68 & 1.93 & 1.71 & 1.84 & -\\
 L254116 & - & 1.59 & 1.75 & 1.60 & 1.62 & -\\
 L254865 & - & 1.84 & 1.97 & 1.91 & 1.79 & -\\
 L254871 & - & 1.86 & 1.95 & 1.80 & 1.84 & - \\ 
 \noalign{\smallskip}
 \hline  
 \noalign{\smallskip}
 Average & 1.59 & 1.65 & 1.84 & 1.73 & 1.77 & 1.82 \\ 
  \noalign{\smallskip}
 \hline 
 \noalign{\smallskip}
 PCC & - & 0.37 & 0.02 & 0.39 & 0.66 & 0.53 \\ 
 P-value & - & 0.33 & 0.95 & 0.23 & 0.03 & 0.47 \\ 
  \noalign{\smallskip}
 \hline 
 \noalign{\smallskip}
\end{tabular}
\vspace{1.5mm}
\end{table*}

\begin{table*}
\centering
\begin{tabular}{l r r r r r r} 
\multicolumn{7}{c}{\textbf{Diffractive scale $r_\text{diff}$ [km]}}\\
 \noalign{\smallskip}
\hline
 \noalign{\smallskip}

Observation ID & LST21-00 & LST00-03 & LST03-06 & LST06-09 & LST09-12 & LST12-15\\ \noalign{\smallskip}
\hline 
\noalign{\smallskip}

L80847 & - & 19.65 & 14.36 & 14.32 & 17.29 & - \\
L80850 & - & 23.80 & 29.94 & 13.46 & 16.58 & - \\
L86762 & - & - & 17.20 & 5.65 & 4.47 & 8.65 \\
L90490 & - & - & - & 14.61 & 9.60 & 21.73 \\
L196421 & - & 13.19 & 22.37 & 16.74 & 33.95 & - \\
L203277 & - & - & - & 7.06 & 3.42 & 4.32 \\
L205861 & - & - & - & 2.88 & 15.95 & 18.46 \\
L246297 & 6.15 & 5.48 & 6.10 & - & - & - \\
L246309 & 11.45 & 39.09 & 26.00 & - & - & - \\
L253987 & - & 20.17 & 6.57 & 3.14 & 3.49 & - \\
L254116 & - & 20.59 & 7.16 & 11.72 & 29.24 & - \\
L254865 & - & 7.30 & 8.54 & 10.10 & 19.85 & - \\
L254871 & - & 14.22 & 11.4 & 6.49 & 22.54 & - \\ 
 \noalign{\smallskip}
\hline  \noalign{\smallskip}
Average & 8.80 & 18.17 & 14.96 & 9.65 & 16.03 & 13.29 \\ 
  \noalign{\smallskip}
\hline \noalign{\smallskip}
PCC & - & 0.02 & -0.19 & 0.32 & -0.71 & -0.01 \\ 
P-value & - & 0.96 & 0.60 & 0.34 & 0.01 & 0.99 \\ 
 \noalign{\smallskip}
\hline
 \noalign{\smallskip}

\end{tabular}
\vspace{1.5mm}
\end{table*}

\begin{table*}
\centering
\begin{tabular}{l r r r r r r} 
\multicolumn{7}{c}{\textbf{Coefficient of determination $\text{R}^2$}}\\ 
 \noalign{\smallskip}
\hline 
\noalign{\smallskip}
Observation ID & LST21-00 & LST00-03 & LST03-06 & LST06-09 & LST09-12 & LST12-15 \\ 
\noalign{\smallskip}
\hline 
 \noalign{\smallskip}
L80847 & - & 0.97 & 0.97 & 0.99 & 0.99 & - \\
L80850 & - & 0.98 & 0.99 & 0.99 & 0.99 & - \\
L86762 & - & - & 0.97 & 0.94 & 0.94 & 0.99 \\
L90490 & - & - & - & 0.98 & 0.98 & 0.99 \\
L196421 & - & 0.99 & 0.89 & 1.00 & 0.98 & - \\
L203277 & - & - & - & 0.93 & 0.89 & 0.96 \\
L205861 & - & - & - & 0.88 & 0.93 & 0.92 \\
L246297 & 0.97 & 0.88 & 0.96 & - & - & - \\
L246309 & 0.92 & 0.94 & 0.95 & - & - & - \\
L253987 & - & 0.98 & 0.96 & 0.90 & 0.96 & - \\
L254116 & - & 0.95 & 0.87 & 0.90 & 0.95 & - \\
L254865 & - & 0.97 & 0.99 & 0.96 & 1.00 & - \\
L254871 & - & 0.98 & 0.99 & 0.94 & 0.97 & - \\ 
 \noalign{\smallskip}
\hline  \noalign{\smallskip}
\end{tabular}
\label{table:ionosphere}
\end{table*}

\begin{figure*}
\centering
\textbf{\\Structure function fits LST00-03\\}
{\includegraphics[width=400px]{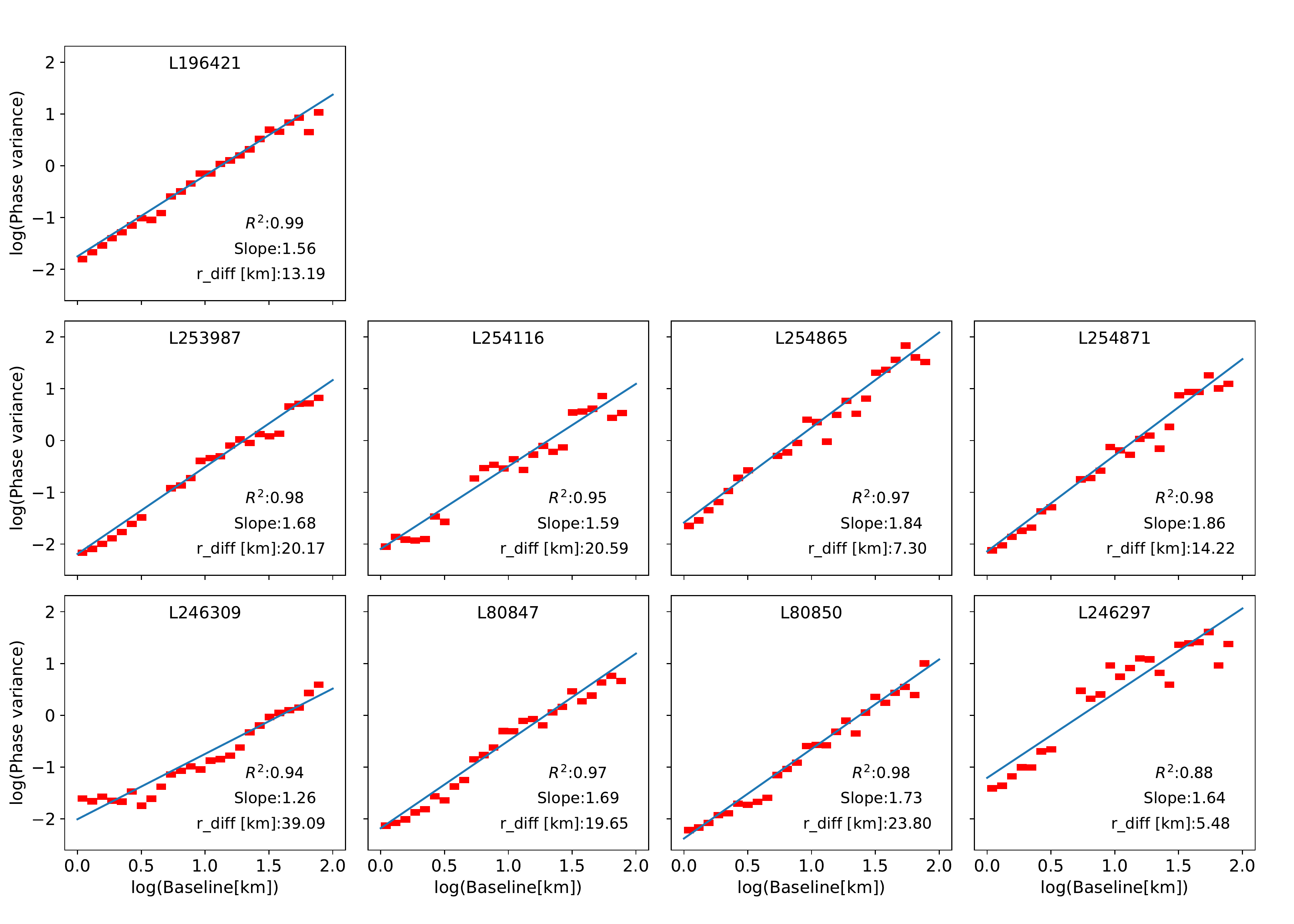}}  \\
\textbf{\\Structure function fits LST03-06\\}
{\includegraphics[width=400px]{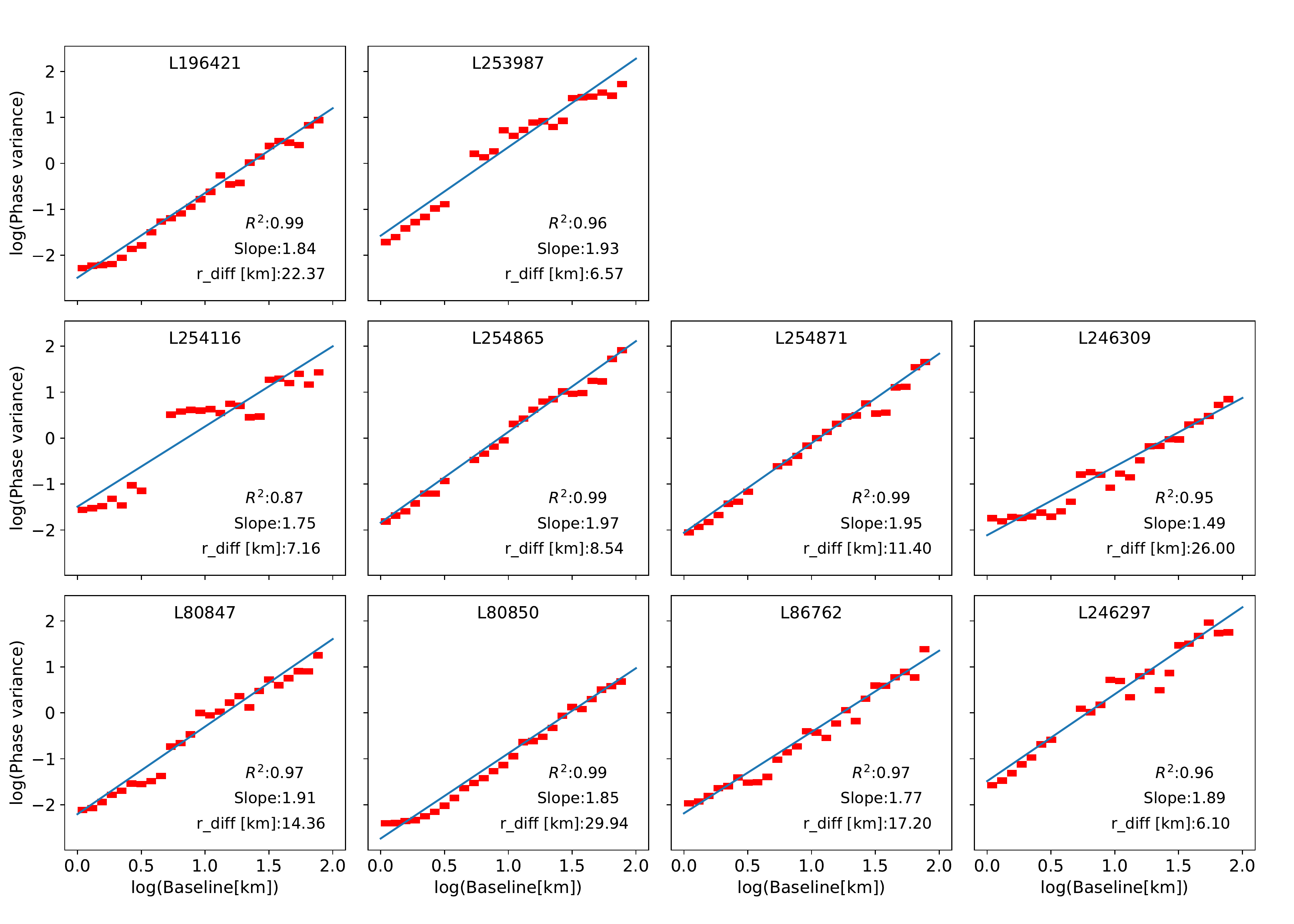}}  \\
\end{figure*}

\begin{figure*}
\centering
\textbf{\\Structure function fits LST06-09\\}
{\includegraphics[width=400px]{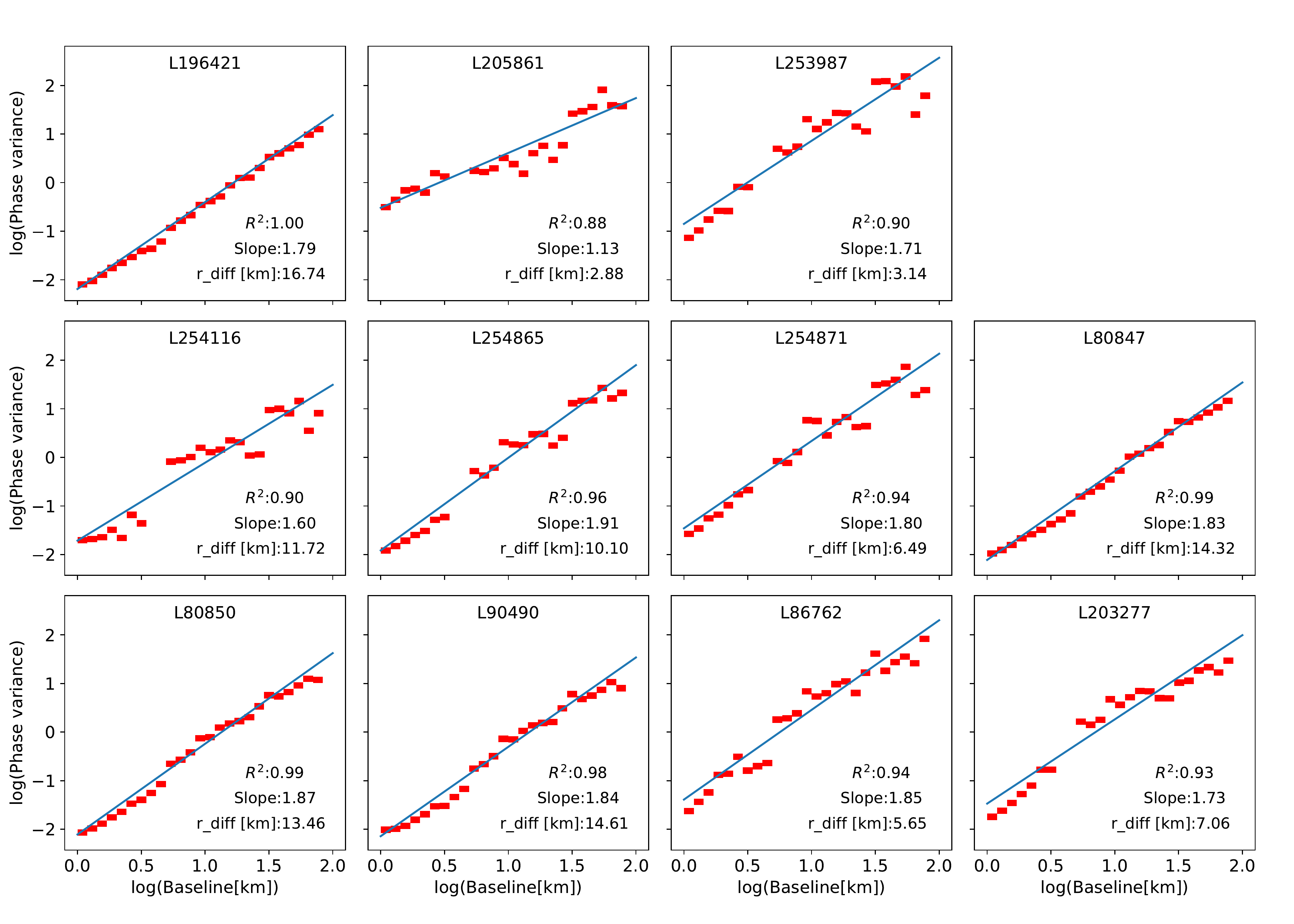}}  \\
\textbf{\\Structure function fits LST09-12\\}
{\includegraphics[width=400px]{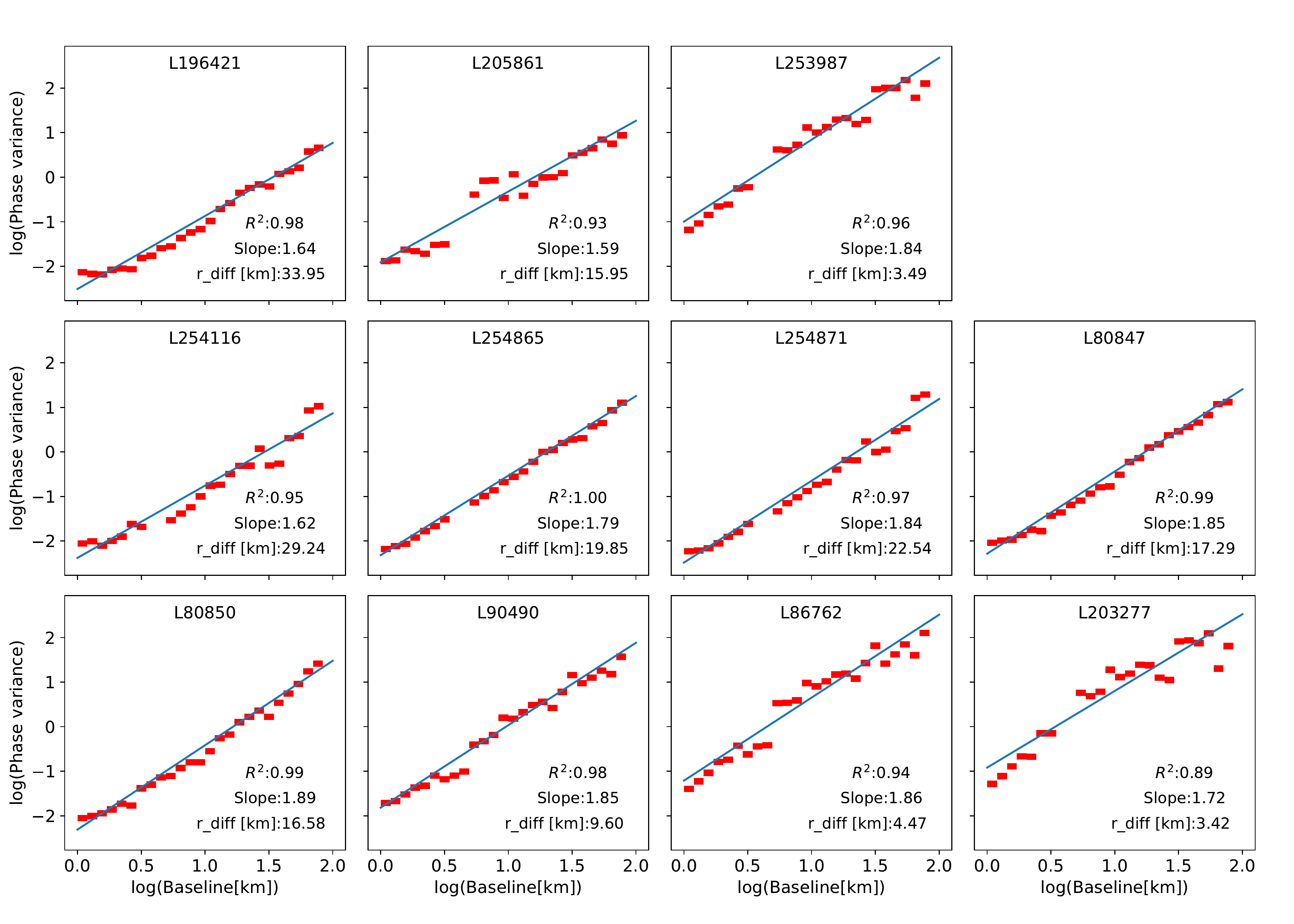}}  \\
\end{figure*}

\end{appendix}
\end{document}